\documentclass[notitlepage,prd,twocolumn,showpacs,preprintnumbers,nofootinbib,superscriptaddress,floatfix,tightenlines]{revtex4-1}
\usepackage{mathrsfs}
\usepackage{amssymb}
\usepackage{amsmath}
\usepackage{graphicx}
\usepackage{hyperref}
\usepackage{amsfonts,amsbsy}
\usepackage{color}

\newcommand{\mbf}{\mathbf}
\newcommand{\xt}{\mathbf{x}_T}

\begin{document}

\title{Nonequilibrium fixed points in longitudinally expanding scalar theories: infrared cascade, Bose condensation and a challenge for kinetic theory}

\author{J.~Berges}
\affiliation{Institut f\"{u}r Theoretische Physik, Universit\"{a}t Heidelberg, Philosophenweg 16, 69120 Heidelberg, Germany}
\affiliation{ExtreMe Matter Institute (EMMI), GSI Helmholtzzentrum f\"ur Schwerionenforschung GmbH, Planckstra\ss e~1, 64291~Darmstadt, Germany}

\author{K.~Boguslavski}
\affiliation{Institut f\"{u}r Theoretische Physik, Universit\"{a}t Heidelberg, Philosophenweg 16, 69120 Heidelberg, Germany}

\author{S.~Schlichting}
\affiliation{Brookhaven National Laboratory, Physics Department, Bldg. 510A, Upton, NY 11973, USA}

\author{R.~Venugopalan}
\affiliation{Brookhaven National Laboratory, Physics Department, Bldg. 510A, Upton, NY 11973, USA}

\begin{abstract}
In Phys.\ Rev.\ Lett.\ 114 (2015) 6, 061601, we reported on a new universality class for longitudinally expanding systems, encompassing strongly correlated non-Abelian plasmas and $N$-component self-interacting scalar field theories. Using classical-statistical methods, we showed that these systems share the same self-similar scaling properties for a wide range of momenta in a limit where particles are weakly coupled but their occupancy is high. Here we significantly expand on our previous work and delineate two further self-similar regimes. One of these occurs in the deep infrared (IR) regime of very high occupancies, where the nonequilibrium dynamics leads to the formation of a Bose-Einstein condensate. The universal IR scaling exponents and the spectral index characterizing the isotropic IR distributions are described by an effective theory derived from a systematic large-$N$ expansion at next-to-leading order. Remarkably, this effective theory can be cast as a vertex-resummed kinetic theory. The other novel self-similar regime occurs close to the hard physical scale of the theory, and sets in only at later times. We argue that the important role of the infrared dynamics ensures that key features of our results for scalar and gauge theories cannot be reproduced consistently in conventional kinetic theory frameworks. 
\end{abstract}
\pacs{11.10.Wx, 12.38.Mh, 67.85.De, 98.80.Cq}

\maketitle

\section{Motivation and overview}

{\it Universality far from equilibrium.} 
In a recent letter we demonstrated universal behavior, in a wide range of momenta, for self-interacting scalars and non-Abelian plasmas when they are longitudinally expanding~\cite{Berges:2014bba}. This is remarkable because the scattering processes controlling the dynamics of scalar and gauge field theories are in principle very different from one another. The universal behavior is observed for systems with sufficiently high occupancy $f \gg 1$ in the limit of weak coupling $\lambda \ll 1$. Here $\lambda$ denotes the quartic self-interaction for scalar fields; for non-Abelian plasmas, it is replaced by the gauge coupling $\sim \alpha_s$. We showed that the far-from-equilibrium dynamics of the different systems in the high occupancy limit is governed by a common nonthermal fixed point with the same dynamical scaling exponents that control the time evolution of characteristic longitudinal and transverse momentum scales. Even detailed properties of the different systems, such as the single particle distributions in the given momentum range, turn out to have identical profiles. This opens up the possibility of studying important aspects of the dynamics of such complex systems using the simplest representative field theory of the corresponding universality class in a well controlled weak-coupling limit. 

{\it Classical-statistical field theory.} A major motivation for these studies are experiments colliding heavy nuclei in the ultrarelativistic high energy limit. In the early stages of such a collision, highly occupied gluon fields with typical momentum
$Q$ are expected to be formed~\cite{McLerran:1993ni,McLerran:1993ka,Iancu:2003xm,Kovner:1995ja,Kovner:1995ts,Krasnitz:1998ns,Krasnitz:1999wc,Krasnitz:2000gz,Krasnitz:2001qu,Krasnitz:2002mn,Lappi:2003bi,Lappi:2006hq,Lappi:2006fp,Lappi:2009xa,Lappi:2011ju}. Since the typical occupancies $\sim 1/\alpha_s(Q)$ are large, the gauge fields can be strongly correlated even for small gauge coupling $\alpha_s(Q)$. In such highly occupied systems dynamical quantum corrections are suppressed at weak coupling and the nonequilibrium quantum dynamics in this transient regime can be accurately mapped onto a classical-statistical problem\footnote{In this regime, quantum corrections are suppressed by one order of the coupling constant even in the anisotropically expanding theories.}. The latter can be solved using real time lattice simulation
techniques~\cite{Khlebnikov:1996mc,Tranberg:2003gi,Micha:2002ey,Micha:2004bv,Berges:2013eia,Berges:2013fga,Berges:2008mr,Berges:2012ev,Schlichting:2012es,Kurkela:2012hp,Scheppach:2009wu,Nowak:2010tm,Nowak:2011sk,Nowak:2013juc,Karl:2013kua,Hebenstreit:2013qxa,Hebenstreit:2013baa,Berges:2013oba,Kasper:2014uaa}.

The classical-statistical field description is limited to the early stages of the evolution since the typical occupancies decrease at later times. It is well known that classical dynamics cannot describe the late time evolution towards thermal equilibrium due to the Rayleigh-Jeans divergence nor can it isotropize in the presence of rapid longitudinal expansion.  The latter property has been confirmed in classical-statistical simulations of the Yang-Mills theory in~\cite{Berges:2013eia,Berges:2013fga} and as well as in classical kinetic theory simulations\footnote{Stated differently, isotropization/thermalization from classical simulations introduces spurious UV divergencies. In that event, the results no longer describe the corresponding quantum field theory dynamics~\cite{Berges:2013lsa,Epelbaum:2014yja}.} of gauge~\cite{Kurkela:2015qoa} and scalar theories~\cite{Epelbaum:2015vxa}.

{\it Effective kinetic theory.} Kinetic theory can in principle describe the late time behavior of systems if the typical occupancies are not too large ($f \ll 1/\lambda$) and if their typical momenta are much larger than the in-medium screening scale $m_D$, such that $p^2 \gg m^2_D \sim \lambda \int_p f(p)/p$, where $ \int_p$ denotes the phase space integration. While these conditions for kinetic theory are not fulfilled for $f\sim 1/\lambda$ at early times, both kinetic theory and classical-statistical field theory can in principle give equally valid descriptions for a wide range of typical perturbative occupancies $1 \ll f \ll 1/\lambda$~\cite{Mueller:2002gd,Jeon:2004dh,Berges:2004yj}. 

A number of weak coupling thermalization scenarios of gauge theories, based on kinetic frameworks, have been developed over the years. These differ primarily in how infrared momentum modes below the scale $m_D$ are treated~\cite{Baier:2000sb,Bodeker:2005nv,Mueller:2005un,Kurkela:2011ti,Kurkela:2011ub,Blaizot:2011xf,Berges:2012ks}, leading to very different paths whereby thermalization is achieved. Since classical-statistical field theory accurately captures the physics of the nonperturbative infrared (IR) regime, it was successfully employed  in~\cite{Berges:2013eia,Berges:2013fga} to single out the kinetic description of~\cite{Baier:2000sb} -- henceforth referred to by the acronym BMSS. As we shall discuss shortly, this result is surprising and requires a deeper understanding of the underlying dynamics. 

We should nevertheless first observe that our results suggest that a comprehensive strategy for weak coupling thermalization is to start with classical-statistical field theory at early times and to continue later with an appropriate kinetic theory, passing from one regime to the other in the time window where both descriptions are valid. This program has led to significant progress in our understanding of the thermalization process of expanding non-Abelian plasmas as summarized in~\cite{York:2014wja,Kurkela:2014tea,Kurkela:2015qoa}. The effective kinetic theory description of~\cite{Kurkela:2015qoa} accurately reproduces the classical-statistical results in the expected regime for weak coupling and sufficiently large momenta. This concerns even detailed properties such as the scaling and shape of the single particle distribution in that range. If applied to stronger couplings, the kinetic descriptions show significant deviations from classical-statistical behavior because of the additional quantum corrections~\cite{Kurkela:2015qoa,Epelbaum:2015vxa} that now become of the same order as the classical contributions. Before extending the results to stronger couplings, it is crucial to fully understand the well controlled weak coupling regime.  

{\it Puzzles.} The result from classical-statistical simulations, demonstrating that BMSS provides the correct kinetic description, was unanticipated. This is because a consistent kinetic description of expanding non-Abelian plasmas would suggest that plasma instabilities~\cite{Mrowczynski:1988dz,Arnold:2003rq,Romatschke:2003ms,Romatschke:2004jh,Rebhan:2004ur} should
qualitatively modify the evolution even at late times~\cite{Kurkela:2011ti,Kurkela:2011ub,Mrowczynski:1988dz,Arnold:2004ti,Bodeker:2005nv}. However classical-statistical simulations of \cite{Berges:2013eia,Berges:2013fga} suggest that plasma instabilities play no significant role beyond the very early stages. While a microscopic understanding of the absence of late time instabilities remains an open question, this observation has been incorporated in the kinetic description of~\cite{Kurkela:2015qoa}.

\begin{figure}[tp!]						
 \centering
 \includegraphics[width=0.5\textwidth]{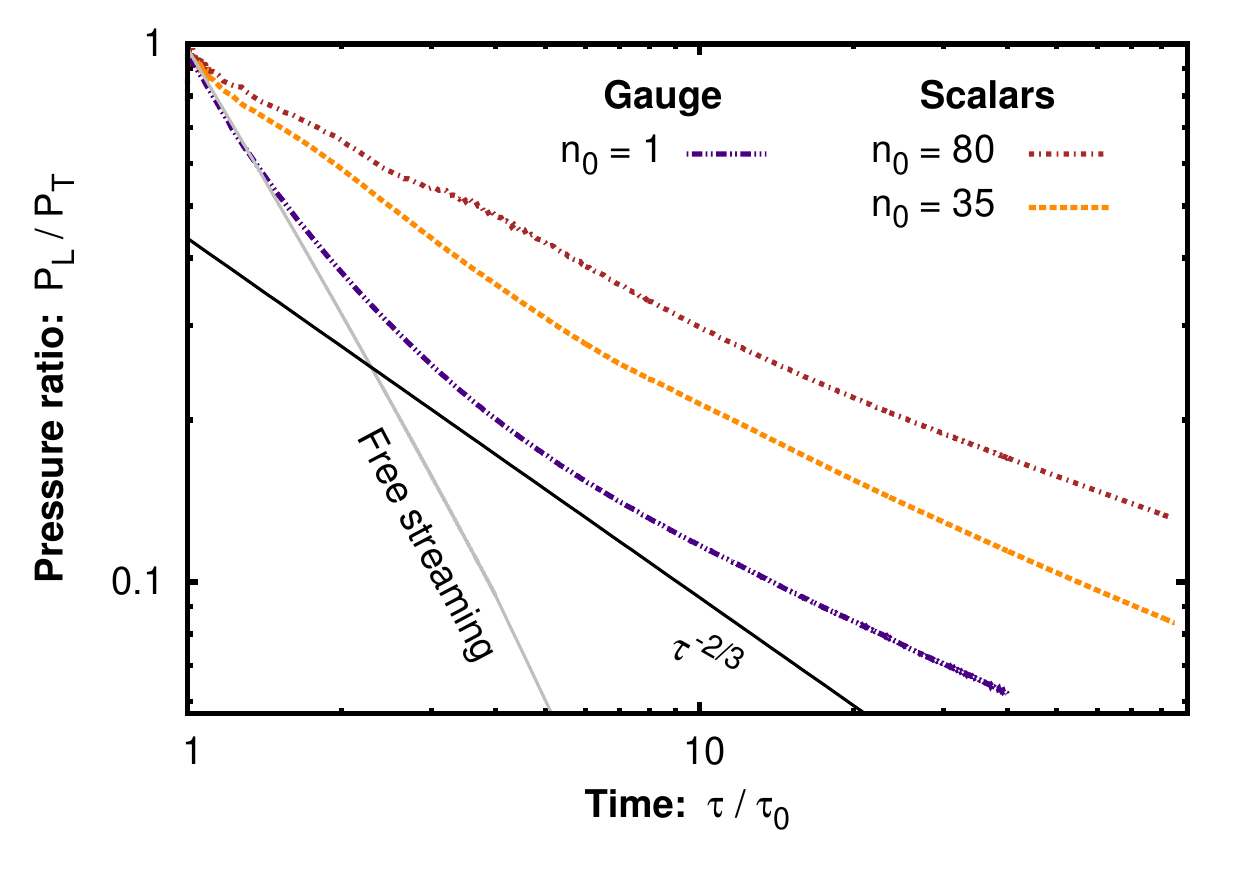}
   \caption{Comparison of the evolution of the longitudinal to transverse pressure ratio in longitudinally expanding scalar and gauge theory. While parametric estimates suggest a scaling as $P_L/P_T \sim (\tau/\tau_{0})^{-2/3}$ in both cases, the observed decay of the longitudinal pressure is significantly slower due to large infrared contributions to the longitudinal pressure. This puzzle is elaborated on further in the text.}
  \label{fig:BulkanisotropyGaugeScalar}
\end{figure}

A further interesting puzzle for longitudinally expanding systems concerns the ratio of the longitudinal pressure to the transverse pressure, $P_L/P_T$, whose value is a measure of the degree of isotropization in the system. In kinetic theory, for non-Abelian gauge theories, small angle scattering dominates for  proper times $ \tau \lesssim Q^{-1} \alpha_s^{-3/2}$. Parametric estimates {\it a la} BMSS would then give 
\begin{equation}
\frac{P_L}{P_T} \,\, \stackrel{\stackrel{\rm kinetic}{\rm theory}}{\sim} \,\, (Q \tau)^{-2/3}\,.
\label{eq:PLPT}
\end{equation} 
However despite the fact that kinetic and classical-statistical descriptions seem to agree well for momenta above the screening scale~\cite{York:2014wja,Kurkela:2015qoa}, the field theoretic simulations of the non-Abelian plasma~\cite{Berges:2013eia,Berges:2013fga} show significant deviations from the kinetic theory scaling of Eq.~(\ref{eq:PLPT}). This is demonstrated in Fig.~\ref{fig:BulkanisotropyGaugeScalar}, where the behavior of the gauge theory~\cite{Berges:2013fga} is shown for isotropic initial conditions.

In contrast to the gauge theory case, the matrix element for self-interacting scalars shows no preference for either small or large angle scattering. Despite this difference, kinetic theory simulations of the expanding scalar theory~\cite{Epelbaum:2015vxa} obtain exactly the result shown in Eq.~(\ref{eq:PLPT}). This agreement with the gauge theory is surprising and not evident from analytical considerations so far. Be that as it may, the classical statistical simulations for the scalar theory shown in Fig.~(\ref{fig:BulkanisotropyGaugeScalar}) deviate significantly from this kinetic theory result for large initial occupancy. 

Since the pressure is a bulk quantity, which sums over all momentum modes, an immediate conclusion from the observed discrepancy is that the infrared regime leads to significant contributions to $P_L/P_T$. The scalar example is useful because the influence of infrared dynamics on the evolution of bulk quantities can be studied without the gauge ambiguities of Yang-Mills theories. 

{\it Infrared dynamics and Bose condensation.}
There are a number of studies suggesting that there are important corrections from infrared modes or Bose condensates at transient stages of the thermalization process for non-Abelian plasmas as well~\cite{Blaizot:2011xf,Blaizot:2013lga,Huang:2013lia,Huang:2014iwa,Scardina:2014gxa,Xu:2014ega,Blaizot:2015wga,Blaizot:2015xga}. However a nonperturbative analysis explaining the classical-statistical simulations results is challenging for the gauge theory. As noted, the situation is very different for scalar field theories. In this case, a detailed analytical understanding can be achieved directly based on the underlying quantum field theory. The  observed universality between longitudinally expanding scalars and non-Abelian plasmas~\cite{Berges:2014bba} in a wide range of momenta could possibly be put to use to obtain further valuable insights into the gauge theory dynamics.

One may also try and simulate infrared effects within kinetic frameworks for scalar field theories. 
One popular kinetic description for scalar field theory, recently implemented for a longitudinally expanding system~\cite{Epelbaum:2015vxa}, approximates the infrared regime by a zero mode that is coupled to a two-to-two scattering kernel. Starting from large initial occupancies, this description gives the scaling Eq.~(\ref{eq:PLPT}) in the weak coupling limit.  Approximating the infrared dynamics by a single zero mode in a kinetic description can be a crude approximation.  
A defect of such an approach, known for systems in a fixed box, is that condensation occurs within a finite time even for infinite volumes~\cite{Semikoz:1994zp,Semikoz:1995rd}. Since any coherent mode cannot spread out faster than the speed of light, this result is an artifact of the approximation. Fortunately, the proper scaling of the condensation time with the volume can be understood in quantum field theory based on systematic large-$N$ expansions~\cite{Berges:2001fi,Aarts:2002dj}. At next-to-leading order (NLO)  the IR dynamics is described by an effective, vertex-resummed kinetic description. This vertex-resummed kinetic description accurately reproduces the results from classical-statistical simulations~\cite{Orioli:2015dxa}, as has been shown before for non-expanding systems.

\begin{figure}[tp!]						
 \centering
 \includegraphics[width=0.5\textwidth]{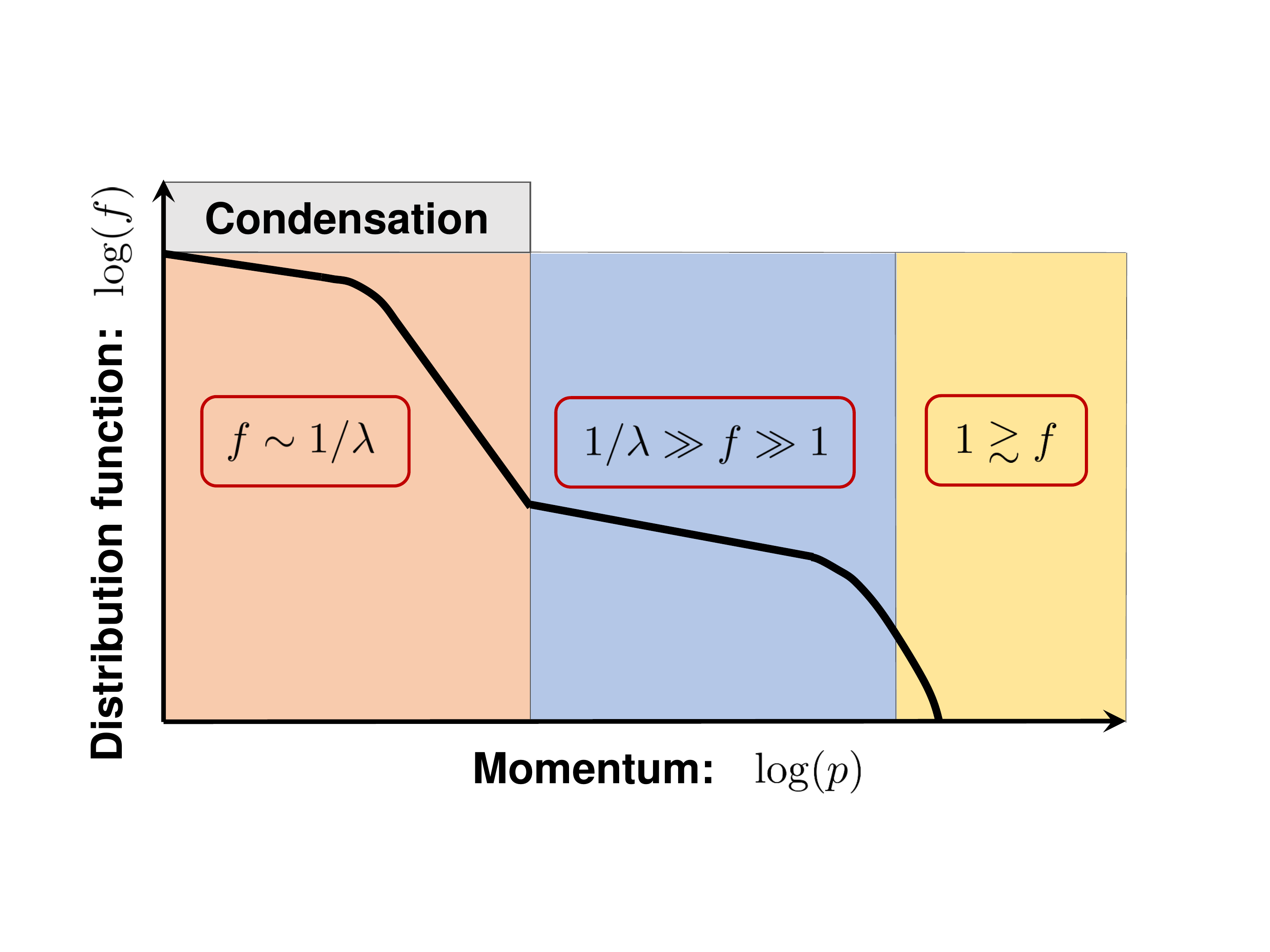}
   \caption{Schematic scaling form of the scalar occupation number distribution $f$ as a function of (transverse) momentum. Classical-statistical field theory is valid for $f \gg 1$. Kinetic theory is restricted to $f \ll 1/\lambda$ if interactions are sufficiently localized in momenta. In scalar field theory, a `vertex-resummed' kinetic description from a $1/N$-expansion at NLO can also describe the IR momentum range.}
  \label{fig:methods}
\end{figure}

{\it Content of this work.} We will analyze longitudinally expanding $N$-component scalar field theories both numerically using classical-statistical simulations and analytically using large-$N$ methods. We investigate the distribution function $f(p)$, whose fixed point scaling form is shown schematically in Fig.~\ref{fig:methods}, with different characteristic momentum ranges. There is a significant range of momenta above a characteristic screening scale, with large but still perturbative occupancies, where $1/\lambda \gg f(p) \gg 1$. This is the regime where comparisons to standard kinetic theory can apply. We note however that such comparisons are predicated on the interactions being sufficiently local in momenta. That this is the case is not assured and understanding whether our results can be explained using kinetic theory is an important objective of our work.

In order to arrive at a deeper understanding of the dynamics sketched in Fig.~\ref{fig:methods}, 
we will expand significantly on our previous work~\cite{Berges:2014bba}. We will study in detail a further scaling regime at hard momenta that was only discussed cursorily in our previous work. The scaling form of the distribution function exhibits an additional infrared regime with nonperturbative occupancies $\sim 1/\lambda$. As noted, this regime is not well described by standard kinetic theory nor in prescriptions assuming an additional single zero mode. Most importantly, it turns out to have significant support in integrals for bulk quantities such as the longitudinal pressure $P_L$. Remarkably, for the scalar field theory, we will show that the effective vertex-resummed kinetic description from  a $1/N$-expansion at NLO describes the physics over the entire IR momentum range.

After specifying the model and parameters in Sec.~\ref{sec:scalars-technical}, we will determine in Sec.~\ref{sec:dynamical-attractor} the universal exponents and scaling functions of the nonthermal fixed point of the longitudinally expanding scalar theory. We show that the system at low momenta behaves as a nonrelativistic superfluid Bose gas with an isotropic distribution function but distinct scaling exponents due to expansion. In particular, we analyze how condensation occurs and find that the condensation time increases with volume. We further compute the properties of the scaling regimes at higher momenta and verify their insensitivity to variations of the initial conditions and microscopic parameters. We discuss the impact of the different momentum ranges on the energy and pressure evolution of the expanding fluid in Sec.~\ref{sec:implications}. We demonstrate that the deviations from standard kinetic descriptions indeed arise from the infrared regime contributing significantly to the longitudinal pressure evolution for large initial occupancies. In Sec.~\ref{sec:large-angle}, we estimate for the  longitudinal pressure the nature of the quantum corrections and the time at which these become relevant for the evolution. A summary is provided in Sec.~\ref{sec:conclusion}. Some technical aspects of the computations are outlined in the Appendices \ref{sec:mode-functions}, \ref{sec:asymptotics} and \ref{sec:collision-integral-low-momenta}.

\section{Scalar field theory with longitudinal expansion}
\label{sec:scalars-technical}

We will consider here the nonequilibrium dynamics of scalar field theories with longitudinal expansion. This is most conveniently done in terms of Bjorken spacetime coordinates using the (longitudinal) proper time $\tau$, longitudinal rapidity $\eta$ and transverse coordinates $\xt=(x^1,x^2)$.  The relation to  Minkowski coordinates $x^0$ and $x^3$ is given by
\begin{align}
\tau = \sqrt{(x^0)^2 - (x^3)^2}\;, \qquad \eta = \textrm{atanh}\left( \frac{x^3}{x^0} \right)\,.
\end{align}
The spacetime metric in Bjorken coordinates takes the form
\begin{align}
g_{\mu \nu}(\tau) = \textrm{diag}(1,-1,-1,-\tau^2)\,,
\label{mat:metric}
\end{align}
which characterizes one dimensional expansion in the longitudinal direction with the expansion rate $1/\tau$.

Within this setup, we will consider a massless $N$-component scalar field theory with $O(N)$ symmetry and quartic interaction defined by the classical action,
\begin{align}
 S[\varphi]=\int d^{4}x \left( \frac{1}{2}(\partial_{\mu}\varphi_a)g^{\mu\nu} (\partial_{\nu}\varphi_a)-\frac{\lambda}{24 N} (\varphi_a\varphi_a)^2 \right).
 \label{mat:action}
\end{align}
Summation over the $a = 1,2,\,\dots\,,N $ scalar field components is implied here and the integration proceeds over $\int d^{4}x \equiv \int d\tau\,d^2\xt\,d\eta\,\sqrt{-g(x)}$ with $\sqrt{-g(x)}=\tau$ resulting from the metric determinant $g(x) = \det(g_{\mu\nu}(x))$. 

In the corresponding quantum theory, the fields are replaced by scalar Heisenberg field operators $\Phi_a(\xt,\eta,\tau)$ and we denote their expectation value by 
\begin{equation}
\phi_a(\tau) = \langle \Phi_a (\xt,\eta,\tau)\rangle
\end{equation}
for spatially homogeneous systems. To make contact with kinetic descriptions, we shall consider the single particle distribution in spatial Fourier space, $f(p_T,p_z,\tau)$. This quantity may be extracted from the connected part of the anticommutator expectation value~\cite{Berges:2004yj} 
\begin{align}
 &F_{ab}(\xt-\xt',\eta-\eta',\tau,\tau') \nonumber \\
 =\,&\frac{1}{2} \langle \{\Phi_a(\xt,\eta,\tau),\Phi_b(\xt',\eta',\tau')\} \rangle - \phi_a(\tau) \phi_b(\tau')\,.
 \label{mat:F-corr-definition}
\end{align}
The distribution function can be defined using the spatial Fourier transform\footnote{The Fourier transform is given by 
\begin{align*}
 F_{ab}(p_T,\nu,\tau,\tau') = \int d^2x_T d\eta\, F_{ab}(\xt,\eta,\tau,\tau')\, e^{-i(\mathbf{p}_T \mathbf{x}_T + \nu \eta)}\,,
\end{align*}
where the rapidity wave number $\nu$ is the conjugate momentum variable to $\eta$.}
of the equal time statistical correlation function $F(p_T,\nu,\tau) \equiv \sum_{a=1}^{N} F_{aa}(p_T,\nu,\tau,\tau)/N$, its temporal derivatives $\ddot{F}(p_T,\nu,\tau) \equiv \tau\, \tau'\, \partial_\tau\, \partial_\tau'\, F(p_T,\nu,\tau,\tau')|_{\tau=\tau'}$ and $\dot{F}(p_T,\nu,\tau)\equiv (\tau\, \partial_\tau + \tau'\, \partial_\tau')\,F(p_T,\nu,\tau,\tau')|_{\tau=\tau'}/2$ according to~\cite{Berges:2004yj}
\begin{align}
 &f(p_T,p_z,\tau) + \frac{1}{2} \nonumber \\
 =\,& \sqrt{F(p_T,\nu,\tau)\ddot{F}(p_T,\nu,\tau)-\dot{F}^2(p_T,\nu,\tau)}\;.
 \label{mat:distr-func-definition}
\end{align}
The longitudinal momentum variable is identified\footnote{This choice of variable is motivated by the form of the longitudinal kinetic term $(\partial_{\eta}^2 \varphi_a)/\tau^2$ in the field equation~(\ref{mat:class-EOM-expanding}) below. In App.~\ref{sec:mode-functions}, we will explain in greater detail how to compute the correlation function numerically and shall discuss an alternative definition of the single particle distribution that is also employed in the literature~\cite{Micha:2004bv,Dusling:2012ig}.} as $p_z \equiv \nu / \tau$.

We employ two different sets of Gaussian initial conditions formulated in terms of the coherent field expectation value $\phi_a(\tau_0)$ and the single particle distribution $f(p_T,p_z,\tau_0)$ at the starting time $\tau_0$. The first set is chosen to exhibit a large characteristic occupancy with
\begin{align}
\label{mat:fluct-IC}
 f(p_T,p_z,\tau_0) = \frac{n_0}{\lambda}\, \Theta\!\left( Q - \sqrt{p_T^2+(\xi_0 p_z)^2} \right)\;,
\end{align}
and vanishing coherent field $\phi_a(\tau_0)=\partial_\tau\phi_a(\tau_0)=0$. Here the parameters describe the initial occupancy $n_0$, initial anisotropy $\xi_{0}$, and the characteristic momentum scale $Q$ of the distribution at initial proper time. We will refer to this as `overoccupation' initial conditions.

Our second set of initial conditions,  are coherent field initial conditions, characterized by a large coherent field, 
\begin{align}
\phi_a(\tau_0) = \sigma_0\, \sqrt{\frac{6 N}{\lambda}}\,\delta_{a1} \;, \quad \partial_\tau\phi_a(\tau_0)=0\;,
 \label{mat:cond-IC}
\end{align}
and vanishing occupancy $f(p_T,p_z,\tau_0)=0$. Details of the numerical implementation of initial conditions are well documented in the literature~\cite{Berges:2012iw,Dusling:2012ig,Berges:2004yj} and are summarized in App.~\ref{sec:mode-functions}.

Both sets of initial conditions admit a rigorous description of the dynamics in terms of classical-statistical field theory to leading order in the coupling $\lambda$~\cite{Son:1996uv,Khlebnikov:1996mc,Aarts:2001yn,Berges:2007ym,Arrizabalaga:2004iw,Mueller:2002gd,Jeon:2004dh}. Since we are interested in the weak coupling limit, with $\lambda \to 0^+$ but $\lambda f$ finite for typical momenta, the mapping is valid at all simulation times. 

For the overoccupation initial conditions (\ref{mat:fluct-IC}), the Monte Carlo sampling and solution of the classical field equation leads to a well defined continuum limit of the corresponding quantum theory~\cite{Aarts:1997kp}. A rigorous mapping of the quantum dynamics with coherent field initial conditions (\ref{mat:cond-IC}) actually involves two steps, which is sometimes overlooked in recent literature. One first  separates the field into a large coherent part and a small fluctuation and linearizes with respect to the latter. Though small initially, the fluctuations grow exponentially because of an instability~\cite{Berges:2012iw,Hatta:2012gq}. Once they become larger than unity, but are still much smaller than the coherent field part, one uses the result as the input for a classical-statistical description which is fully nonlinear. The whole procedure has a well defined continuum limit such that one gets the physical quantum result~\cite{Son:1996uv,Khlebnikov:1996mc,Hatta:2012gq}. In the following, we will employ the coherent field initial conditions of Eq.~(\ref{mat:cond-IC}) for finite values $\lambda \ll 1$ and momentum cutoffs only to show that after the parametric time $\sim \log^{3/2}( \lambda^{-1})$ the initial field decays~\cite{Berges:2012iw} and one recovers the fluctuation initial conditions of Eq.~(\ref{mat:fluct-IC}).

We solve numerically the nonlinear classical field equations of motion\footnote{The coupling dependence can be conveniently scaled out of the equations with $\varphi \rightarrow \varphi/\sqrt{\lambda}$, thus entering only the initial conditions.},
\begin{align}
\left( \partial_\tau^2 + \frac{1}{\tau}\,\partial_\tau - \partial_T^2 - \frac{1}{\tau^2}\partial_{\eta}^2 + \frac{\lambda}{6 N}\varphi^2 \right)\varphi_a = 0\;,
\label{mat:class-EOM-expanding}
\end{align}
using leapfrog integration. We discretize the equations on anisotropic lattices of size $N_T^2 \times N_{\eta}$ with lattice spacings $a_T,\;a_{\eta}$, and use a time step $a_\tau$ which is small compared to all other dimensional parameters, namely, $a_\tau \ll \tau,a_T, a_{\eta} \tau$, to guarantee the stability of the algorithm. We have explicitly checked that our results are insensitive to changes of the discretization parameters.

\begin{figure}[t]						
 \centering
 \includegraphics[width=0.5\textwidth]{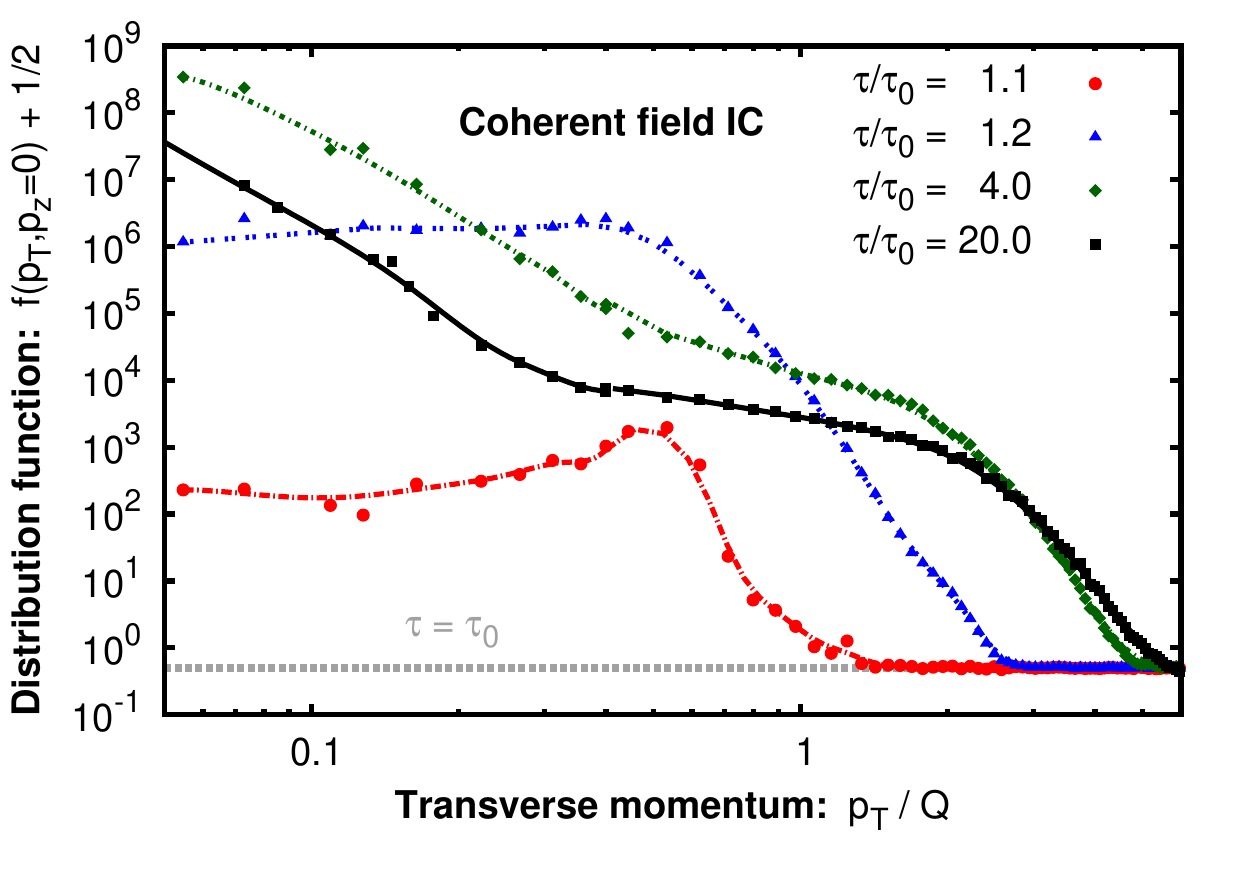}
  \caption{Snapshots of the occupation number distribution at vanishing longitudinal momentum at different times for the coherent field initial conditions in Eq.~(\ref{mat:cond-IC}), with coupling constant chosen to be $\lambda=10^{-4}$. Starting with the pure vacuum (gray) distribution, fluctuations are built up via the parametric resonance mechanism (red circles) and lead to an overoccupied system (blue triangles). The latter approaches a nonthermal fixed point (green rhombi and black squares).}
  \label{fig:coherent-field-IC}
\end{figure}

In Fig.~\ref{fig:coherent-field-IC}, we follow the time evolution for coherent-field initial conditions in Eq.~(\ref{mat:cond-IC}). These initial conditions lead to a parametric resonance instability at early times in the static case~\cite{Kofman:1994rk,Greene:1997fu,Berges:2002cz} as well as the expanding scalar field theory~\cite{Berges:2012iw,Hatta:2012gq} despite the longitudinal expansion. As shown in Fig.~\ref{fig:coherent-field-IC}, this instability results in rapid particle production (red circles) and an overoccupation of low momentum modes built up on a short time scale (blue triangles) while the coherent field decays. In this case, we define the characteristic momentum scale as $Q \equiv \sigma_0$ of Eq.~(\ref{mat:cond-IC}) and identify an overoccupied system with a similar form as given by Eq.~(\ref{mat:fluct-IC}). Hence starting with an overoccupied single particle distribution may be regarded as an already  time evolved system with coherent field initial conditions. At later times (green rhombi and black squares), the system approaches a nonthermal fixed point, whose universal properties we will analyze in the following.

We note that a similar argument can be made in the case of the longitudinally expanding non-Abelian plasma. Starting from the Glasma, which consists of strong boost invariant color fields~\cite{McLerran:1993ni,McLerran:1993ka,Iancu:2003xm,Kovner:1995ja,Kovner:1995ts}, particles are rapidly produced via plasma instabilities~\cite{Romatschke:2005pm,Romatschke:2006nk,Romatschke:2005ag,Fukushima:2011nq,Berges:2012cj} and lead to overoccupation~\cite{Berges:2014yta}. An overoccupied non-Abelian plasma approaches the nonthermal fixed point found in Refs.~\cite{Berges:2013eia,Berges:2013fga} in the weak coupling limit.

\section{Nonthermal fixed points of longitudinally expanding scalars}
\label{sec:dynamical-attractor}

Classical-statistical field theory efficiently describes the far from equilibrium dynamics of highly occupied many-body systems. Numerical simulations in this context reveal that highly occupied systems across a wide range of energy scales spanning inflationary cosmology~\cite{Khlebnikov:1996mc,Micha:2002ey,Micha:2004bv}, heavy ion collisions~\cite{Berges:2013eia,Berges:2013fga,Berges:2008mr,Berges:2012ev,Schlichting:2012es,Kurkela:2012hp}, ultracold atoms~\cite{Scheppach:2009wu,Nowak:2010tm,Nowak:2011sk,Karl:2013kua} and, as has been recently suggested, holographic superfluids~\cite{Ewerz:2014tua} display universal behavior characterized by nonthermal fixed points. 

In the vicinity of a nonthermal fixed point, strongly correlated quantum systems lose their sensitivity to the initial conditions and other microscopic parameters of the theory~\cite{Berges:2013eia,Berges:2013fga,Berges:2013lsa}. The single particle distribution function can then be divided into separate scaling regions in momentum space, each of which follows a self-similar evolution
\begin{equation}
f(p_T,p_z,\tau) = \tau^\alpha f_S \left(\tau^\beta p_T, \tau^\gamma p_z \right)\,,
\label{mat:scaling}
\end{equation}
where $\tau$ is the proper time. Different aspects of the dynamics can be described in terms of the universal scaling exponents $\alpha$, $\beta$ and $\gamma$ and (proper) time independent scaling function $f_S(p_T,p_z)$~\cite{Zakharov:1992,Micha:2004bv,Berges:2014bba}.

In general a nonthermal fixed point can exhibit multiple scaling regions simultaneously, with a different scaling behavior realized in each of these separate momentum regions. Such multiple scaling behavior can emerge as a consequence of multiple conservation laws such as the self-similar transport of a different conserved quantity such as energy or particle number in one regime compared to another. Another possible reason includes a change of the underlying dynamics, for example, from a perturbative to a nonperturbative mechanism below a characteristic momentum scale. In the context of wave turbulence, each momentum region is usually referred to as an \emph{inertial range} of momenta, with various examples provided in the literature \cite{Micha:2004bv,Orioli:2015dxa,Berges:2008wm,Berges:2008sr,Scheppach:2009wu,Nowak:2010tm,Berges:2012us,Nowak:2012gd,Berges:2013lsa}.

Within the classical-statistical simulations of the expanding scalar theory in \cite{Berges:2014bba}, we identified three distinct inertial scaling regimes: i) An inverse particle cascade at very soft momenta with large occupancies $f \gtrsim 1/\lambda$, ii) an intermediate range of momenta extending dynamically up to momenta close to the hard momentum scale $Q$ of the system, and iii) a hard momentum regime for momenta $p_T \gtrsim Q$. In the following, we will expand on the discussion of the longitudinally expanding scalar theory in \cite{Berges:2014bba} and present detailed results uncovering the properties of the three inertial regimes. 

Since the nonthermal fixed point structure is insensitive to the details of the initial conditions, we will first focus on the time evolution for a single initial condition and then verify that the results are insensitive to the initial conditions (summarized in Table \ref{tab:Discretization}) further below for parts of the attractor. Unless stated otherwise, we consider a four component ($O(4)$) scalar theory with overoccupation initial conditions, as in Eq.~(\ref{mat:fluct-IC}), with an initially isotropic system $(\xi_0 = 1)$ and with the occupancy parameter $n_0 = 35$ at the initial time $Q\tau_0 = 10^3$. All quantities in this paper are normalized by appropriate dimensional scales. If not written explicitly, we imply $\tau \mapsto \tau/\tau_0$ and $p_i \mapsto p_i / Q$, which also holds for the scaling behavior introduced in Eq.~(\ref{mat:scaling}).

\begin{table}[b!]
\begin{tabular}{||c|c|c|c||c|c|c|c||}
\hline \hline
\multicolumn{4}{||c||}{Configuration} & \multicolumn{4}{c||}{Lattice parameters} \\
 $\xi_0$ & $n_0$ & $Q\tau_0$ & $N$ & $N_T$ & $N_\eta$ & $Qa_{T}$ & $ a_{\eta}/10^{-5}$ \\ \hline \hline
$0.5$ & $5$ & $1000$ & $4$ & $128$ & $768$ & $1$ & $5$ \\ \hline
$1$ & $4.5$ & $1000$ & $4$ & $128$ & $768$ & $1$ & $5$ \\ \hline
$1$ & $5$ & $1000$ & $4$ & $128-256$ & $512-4096$ & $0.5 - 1$ & $2.5-5$ \\ \hline
$1$ & $6$ & $1000$ & $4$ & $128$ & $768$ & $1$ & $5$ \\ \hline
$1$ & $7.5$ & $1000$ & $4$ & $128$ & $768$ & $1$ & $5$ \\ \hline
$1$ & $15$ & $1000$ & $4$ & $192$ & $768$ & $0.67$ & $5$ \\ \hline
$1$ & $35$ & $1000$ & $4$ & $96-384$ & $384-768$ & $0.25-2.5$ & $4-10$ \\ \hline
$1$ & $80$ & $1000$ & $4$ & $128$ & $768$ & $0.4$ & $5$ \\ \hline
$2$ & $15$ & $1000$ & $4$ & $128$ & $768$ & $0.85$ & $5$ \\ \hline
$4$ & $15$ & $1000$ & $4$ & $128$ & $1024$ & $1$ & $5$ \\ \hline \hline
$1$ & $35$ & $100$ & $4$ & $192$ & $1024$ & $0.75$ & $25$ \\ \hline
$1$ & $5$ & $1000$ & $2$ & $192$ & $768$ & $0.5$ & $5$ \\ \hline
\multicolumn{2}{||c|}{Eq.~(\ref{mat:cond-IC})} & $1000$ & $4$ & $192;256$ & $256-768$ & $0.2 - 0.9$ & $6-20$ \\ \hline
\hline
\end{tabular}
\caption{\label{tab:Discretization}Discretization parameters for different initial conditions. Unless stated otherwise, these parameters are employed for all classical-statistical lattice simulations. In the following, the configurations of the last three lines are referred to as ``$Q\tau_0 = 100$", ``$\mathcal{O}(2)$ scalars" and ``Coherent field IC" (c.f. Eq.~(\ref{mat:cond-IC})), respectively.}
\end{table}

\begin{figure}[tp!]						
 \centering
 \includegraphics[width=0.5\textwidth]{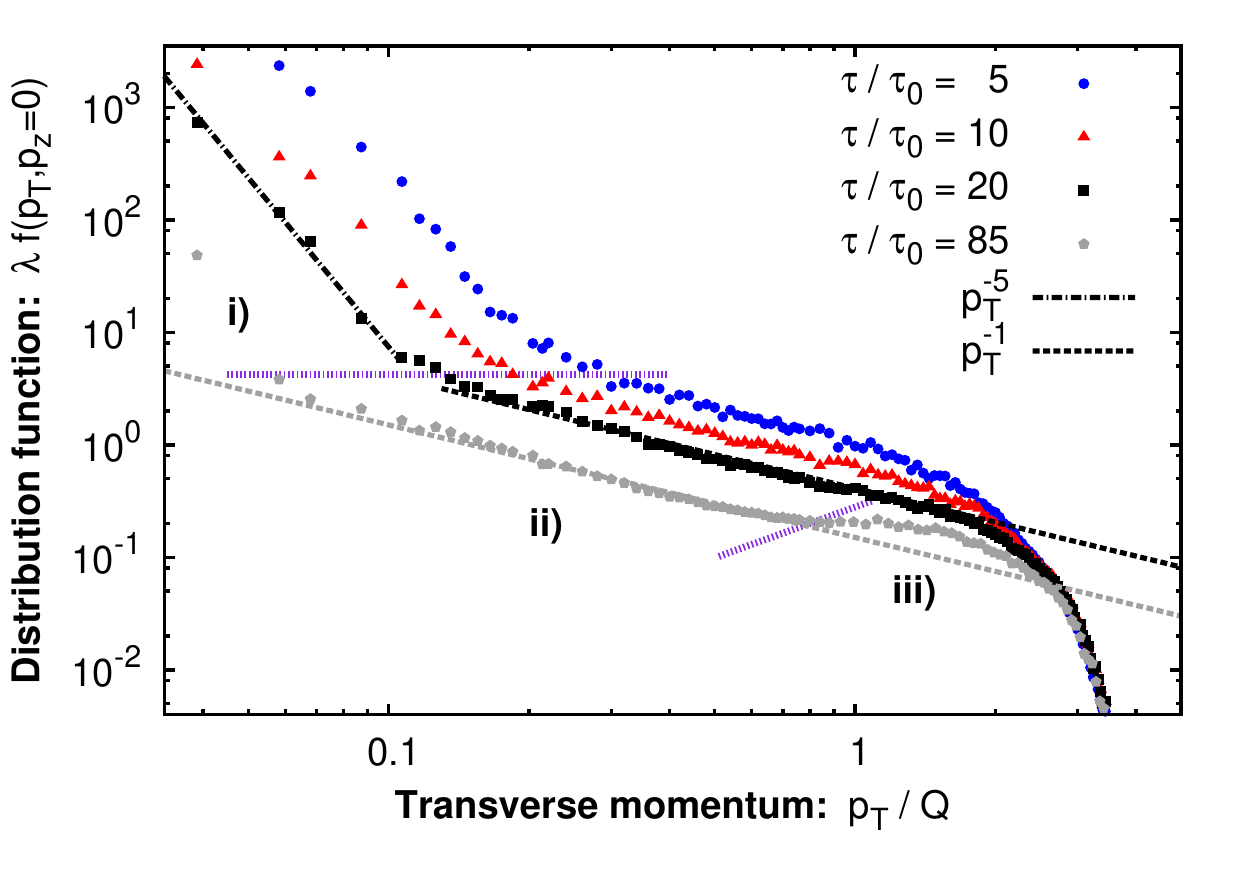}
  \caption{Dependence of the single particle distribution on the transverse momentum (for vanishing longitudinal momentum $p_z=0$) at various times. One observes the emergence of three distinct scaling regimes labeled as i), ii) and iii) and separated by purple dashed lines. Each regime corresponds to the inertial range of momenta for a nonthermal fixed point as discussed in detail below. The black and gray dashed lines illustrate the power law dependence in regimes i) and ii). }
  \label{fig:spectrum-pz0-BoxIC}
\end{figure}

Our results for the transverse momentum dependence of the spectrum at different proper times in the evolution are compactly summarized in Fig.~\ref{fig:spectrum-pz0-BoxIC}, which shows snapshots of the single particle distribution $f(p_T,p_z=0,\tau)$ for vanishing longitudinal momenta. One clearly observes the emergence of three distinct scaling regimes, at small, intermediate and high transverse momenta. We reported in \cite{Berges:2014bba} that the low momentum behavior -- indicated as i) -- is characterized by a strong infrared enhancement with an approximate $p_T^{-5}$ power law dependence while the intermediate momentum regime -- indicated as ii) -- features an approximate $p_T^{-1}$ power law shown by the gray and black dashed lines. 

The inertial ranges of momenta for the scaling regimes i) and ii) are seen to shift dynamically towards lower momenta as a function of time as indicated by the purple dashed lines in Fig.~\ref{fig:spectrum-pz0-BoxIC}. At late times, an additional scaling window emerges at momenta $p_T\gtrsim Q$. This novel regime was unanticipated and we shall discuss it at length shortly. In this regime -- indicated as iii) --  the single particle distribution becomes approximately independent of the transverse momentum up to very high transverse momenta where the spectrum shows a rapid fall off.

Each of the three momentum regions corresponds to a different self-similar evolution of the single particle distribution with the scaling behavior in Eq.~(\ref{mat:scaling}). The exponents determine the evolution of the typical amplitude $f_{\text{typ}} \sim \tau^\alpha$ and of typical momenta $p_{T,\text{typ}} \sim \tau^{-\beta}$ and $p_{z,\text{typ}} \sim \tau^{-\gamma}$. Moreover the signs of $\beta$ and $\gamma$ provide the effective direction of transport of particles or energy within the inertial range of momenta. We will extract the universal exponents and the scaling function of each scaling region in subsequent sections.




\begin{figure}[t]						
 \centering
 \includegraphics[width=0.5\textwidth]{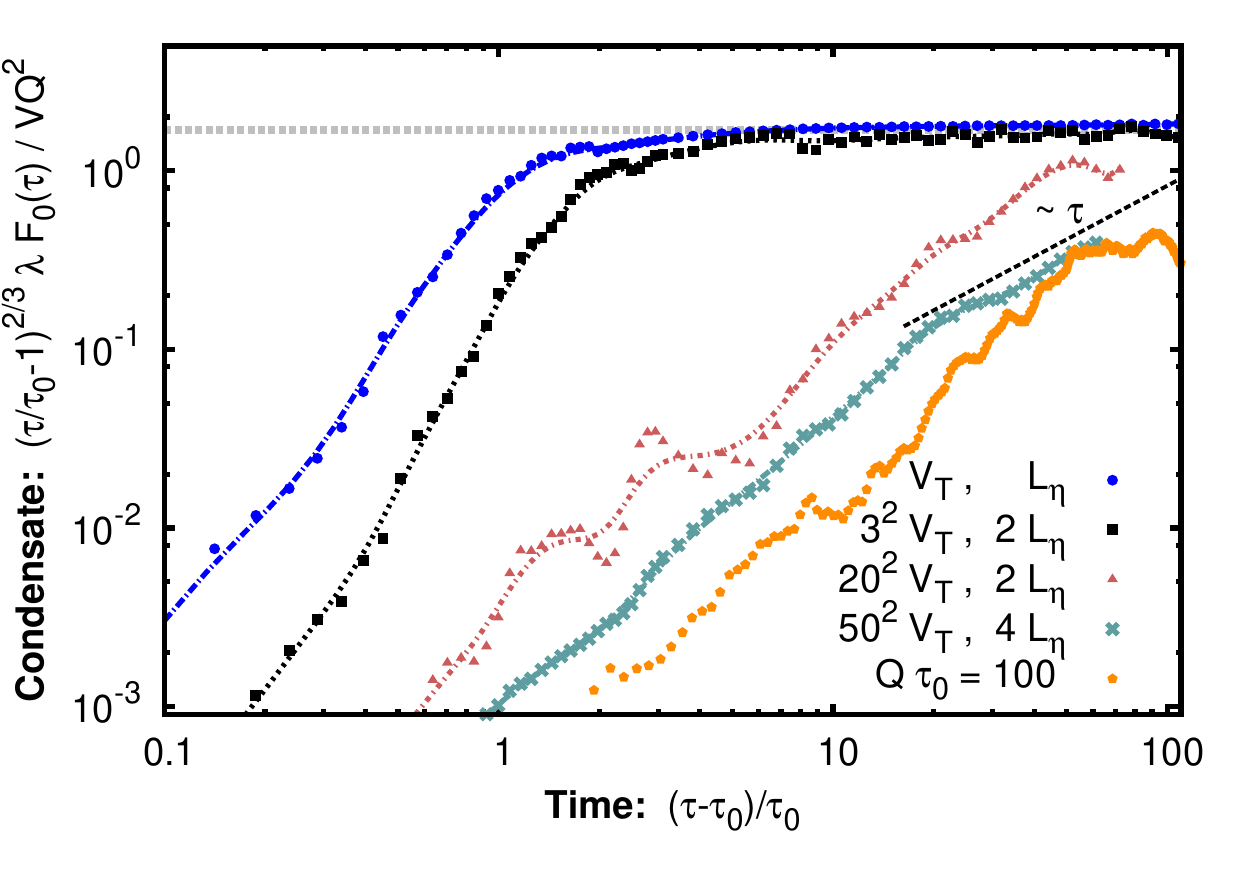}
  \caption{ Time evolution of the condensate observable with overoccupation initial conditions, plotted for different volumes $V \equiv V_T L_\eta$. The lattice spacings for the smallest volume are $Q a_T = 0.75$ and $a_\eta = 5 \cdot 10^{-5}$ on a $64^2 \times 384$ grid. At sufficiently late times, the results become independent of the volume, signaling the formation of a condensate. The condensation time increases with volume. For large volumes, the condensate growth follows power laws that are marked by a dashed black line. A similar late-time behavior is found for other initial parameters, as for an earlier initial time $Q\tau_0 = 100$.}
  \label{fig:condensation}
\end{figure}

\subsection{Infrared cascade and Bose condensation}
\label{sec:low-momenta}

We will first present a detailed analysis of the inertial range at low momenta, marked by i) in Fig.~\ref{fig:spectrum-pz0-BoxIC} with an approximate $p_T^{-5}$ power law in the transverse momentum spectrum. In the study of scalar field theories with static geometry, such a low momentum region is associated with an inverse particle number cascade of an effectively massive theory and leads to the formation of a Bose condensate~\cite{Berges:2012us,Orioli:2015dxa}. 

We can similarly investigate the presence of a dynamically generated Bose condensate in the longitudinally expanding system by studying the finite volume scaling of correlation functions of the scalar field. In an approach similar to that considered previously for the non-expanding theory, we examine the properties of the statistical correlation function $F_0(\tau) \equiv F(p_T=0,\nu=0,\tau)$ of Eq.~(\ref{mat:F-corr-definition}). While the fluctuations at finite momenta are essential for the formation of the condensate via an inverse particle cascade, a volume independent correlation function at zero momentum $F_0(\tau)/(V_T L_\eta) = \phi_0^2(\tau)$ signals the existence of a Bose condensate. Our results for this condensate observable are shown in Fig.~\ref{fig:condensation}, where we compare the time evolution for different volumes. Starting from overoccupation initial conditions, one observes a rapid rise of the zero momentum mode at early times. Eventually the condensate observable becomes volume independent thereby signaling the formation of a condensate. As we noted in our previous letter~\cite{Berges:2014bba}, the condensate then decreases as $F_0(\tau) \sim \tau^{-2/3}$. 

The condensation time $\tau_c$ can be viewed as the onset time of volume independent evolution.  In our case, the rescaled zero mode $(\tau-\tau_0)^{2/3} F_0$ becomes constant in time (as marked by the constant gray dashed line). While for smaller volumes the condensate is created during the simulation time, for larger volumes its formation is not completed. Nevertheless from the ordering of the curves one observes that the condensation time increases with volume. While conventional kinetic descriptions have a finite condensation time even for infinite volumes~\cite{Semikoz:1994zp,Semikoz:1995rd}, the fact that the condensation time diverges with increasing volume is expected from causality and is an important feature of our nonperturbative numerical approach. 

We note that for the longitudinally expanding system causality also forbids the formation of a Bose-Einstein condensate if the rapidity range is too large. This is because at any given time $\tau$, two points which are separated by a larger rapidity range than $\Delta \eta_{c} = 2 \ln(\tau_c/\tau_0)$ in the longitudinal direction are not causally connected. Nevertheless one may still observe the emergence of long range correlations for a system with sufficiently small rapidity extent $\Delta \eta\leq \Delta \eta_c$ and all our considerations are only valid in the rapidity range where we do observe condensation.

For nonexpanding systems, both relativistic and nonrelativistic lattice simulations also indicate an increasing condensation time with volume, as $t_c \sim V^{2/3}$~\cite{Orioli:2015dxa}. This results from the condensation behavior $F_0 \sim t^{3/2}$, a consequence of the inverse particle cascade from the low momentum region that populates the zero mode~\cite{Svistunov:1991,Kagan:1992,Berloff:2002,Berges:2012us,Nowak:2012gd}.

\begin{figure}[tp!]						
 \centering
 \includegraphics[width=0.5\textwidth]{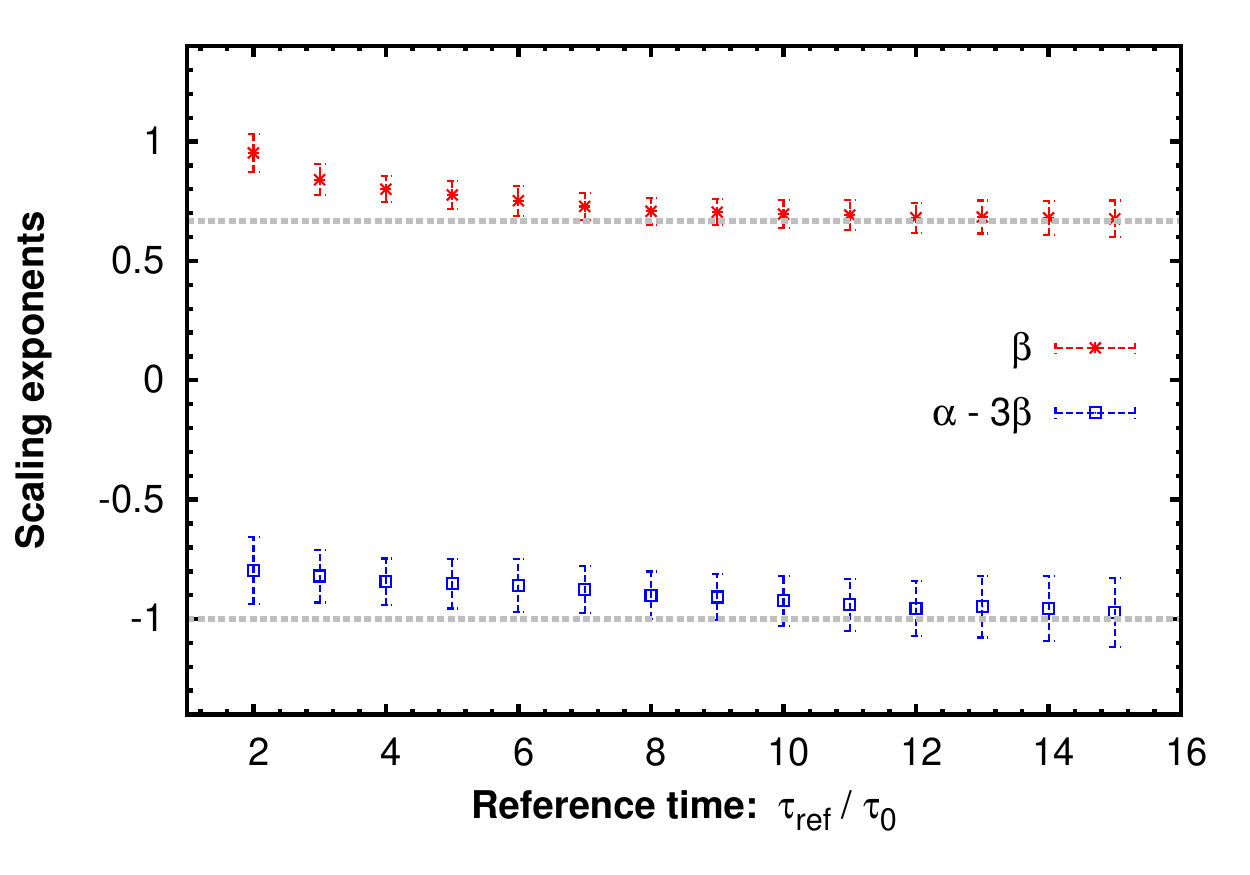}
 \caption{The exponents $\beta$ and $\alpha - 3\beta$ extracted at different reference times $\tau_{\text{ref}}$. The gray dashed lines correspond to the values $2/3$ and $-1$ extracted from a scaling analysis within a vertex-resummed kinetic description.}
  \label{fig:scaling-low-momenta-evolution}
\end{figure}

As can be seen in Fig.~\ref{fig:condensation} for larger volumes, the condensate observable in the expanding theory $\tau^{2/3} F_0(\tau)/V$ also follows an approximate power law behavior in time. We expect from analytic considerations detailed in Sec.~\ref{sec:infrared-explanation} that the  power law exponent approaches unity at late times. This leads to an increasing condensation time with volume $\tau_c \sim V$, as we will discuss in detail in Sec.~\ref{sec:infrared-explanation}. Indeed the results of Fig.~\ref{fig:condensation} seem to support this behavior although with sizable discrepancies at earlier times and for smaller volumes.\footnote{Discrepancies from $\Delta \tau^{2/3} F_0(\tau)/V \sim \Delta \tau$ at earlier times with $\Delta \tau = \tau-\tau_0$ are expected due to finite time effects when the system is not yet close enough to the nonthermal fixed point. Such early time effects can also be observed in the scaling behavior of other quantities as shown explicitly in Fig.~\ref{fig:scaling-low-momenta-evolution}. Similarly, simulations with small volumes do not show the power law behavior at all because the low momentum region is not completely formed and the system is thus too far from the attractor on the time scales of condensation.}. A similar power law evolution can also be observed in simulations with different initial parameters and sufficiently large volume, as seen for the system with the earlier initial time $Q \tau_0 = 100$ for comparison.

\begin{figure}[tp!]						
 \centering
 \includegraphics[width=0.5\textwidth]{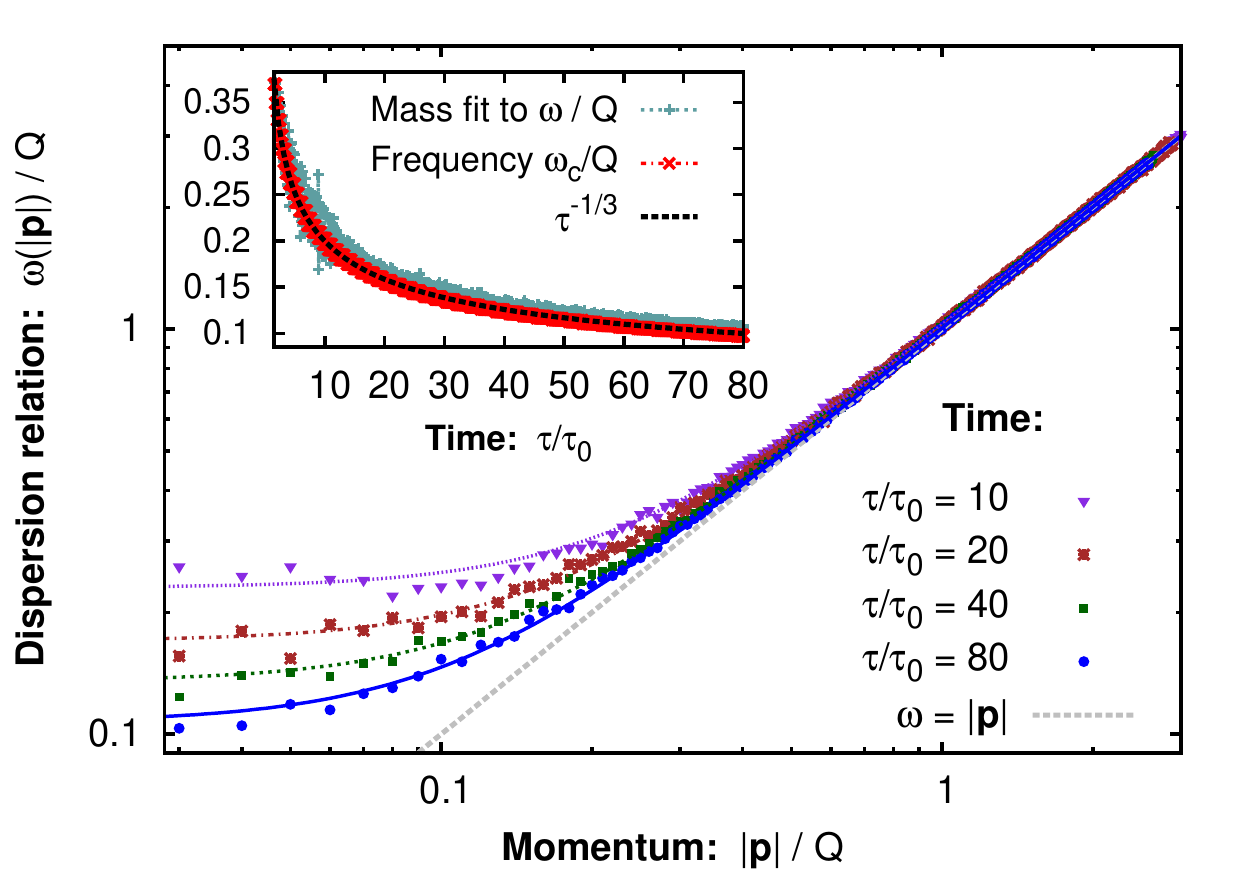}
  \caption{Effective dispersion relation at different times. Deviations from the $\omega=|\mathbf{p}|$ behavior show that a mass gap is generated for low momentum excitations. The time dependence of the effective mass is shown in the inset, extracted from the zero momentum frequency via $\sqrt{m^2 + |\mathbf{p}|^2}$ fits to the dispersion relation and by measuring the oscillation frequency of the zero momentum mode. The effective mass decreases according to a $\tau^{-1/3}$ power law.}
  \label{fig:dispersion-relation}
\end{figure}

In the scalar field theory for a static box, both the condensate and a dynamically generated mass were observed in lattice simulations~\cite{Orioli:2015dxa}. We have also checked the existence of a dynamically generated mass for the longitudinally expanding case. In analogy to the static box  case~\cite{Berges:2004yj,Orioli:2015dxa}, we define 
\begin{align}
 \omega(p_T,p_z,\tau) = \sqrt{\frac{\ddot{F}(p_T,\nu,\tau)}{\tau^2\, F(p_T,\nu,\tau)}}
 \label{mat:dispersion-relation}
\end{align}
to estimate the effective dispersion relation in the medium. We find that up to statistical fluctuations, the dispersion relation is independent of the polar angle $\theta = \arctan (p_T/p_z)$ in the entire momentum space and only depends on the absolute momentum $|\mathbf{p}| = \sqrt{p_T^2 + p_z^2}$.

Our results for $\omega(|\mathbf{p}|,\tau)$, averaged over the angle $\theta$, are shown in Fig.~\ref{fig:dispersion-relation} at different times. A comparison with the (dashed) lines shows that the effective dispersion relation is well described by a relativistic quasi-particle dispersion
\begin{align}
\omega(p_T,p_z,\tau)=\sqrt{m^2(\tau) + |\mathbf{p}|^2}\;,
\end{align}
with a time dependent effective mass $m(\tau)$. 

The time dependence of $m(\tau)$ is shown in the inset of Fig.~\ref{fig:dispersion-relation}, where we compare $m(\tau)$ extracted from the dispersion relation to the one extracted from the oscillation frequency $\omega_{c}(\tau)$ of the unequal-time correlator at zero momentum $F(p_T=0,\nu=0,\tau,\tau_0)$. Both measurements agree and show that the dynamically generated mass decreases as a function of time according to 
\begin{align}
 m(\tau) \sim \tau^{-1/3}\;.
 \label{mat:mass-evolution}
\end{align}

Because of the large occupancy at low momenta (region i) in Fig.~\ref{fig:spectrum-pz0-BoxIC}) the mass is expected to be dominated by modes in the infrared scaling region. Using $m^2 \sim \int d^3p\,f/\omega$ to estimate the evolution of the mass, one has for infrared modes
\begin{align}
 m^2 \sim \frac{n}{m}\,.
 \label{mat:mass-equation}
\end{align}
Particle number conservation $n \sim \tau^{-1}$ (which will be demonstrated later in this section) leads to $m \sim \tau^{-1/3}$ as observed in (\ref{mat:mass-evolution}). To arrive at Eq.~(\ref{mat:mass-equation}) we used $\omega(p) \approx m$ which is true for the infrared modes with $|\mathbf{p}| \lesssim m$. The contribution of the condensate to the effective mass scales in the same way. In contrast, one can estimate the contribution from the harder momentum regions ii) and iii) in Fig.~\ref{fig:spectrum-pz0-BoxIC} as $m^2_{\text{Hard}} \sim \int d^3p\, \lambda f/\omega \sim \tau^{-1}$, which is subleading because of its faster decay in time.

We note that this is quite different from thermal equilibrium where the effective mass is dominated by hard momenta at the temperature scale. The difference can be understood as follows.  Close to the nonthermal fixed point, the low momentum region is enhanced and its contribution to the mass decreases slower than at hard momenta, dominating at late times. In contrast, no such enhancement at low momenta exists in thermal equilibrium and all contributions to the mass are stationary.

\begin{figure}[tp!]						
 \centering
 \includegraphics[width=0.5\textwidth]{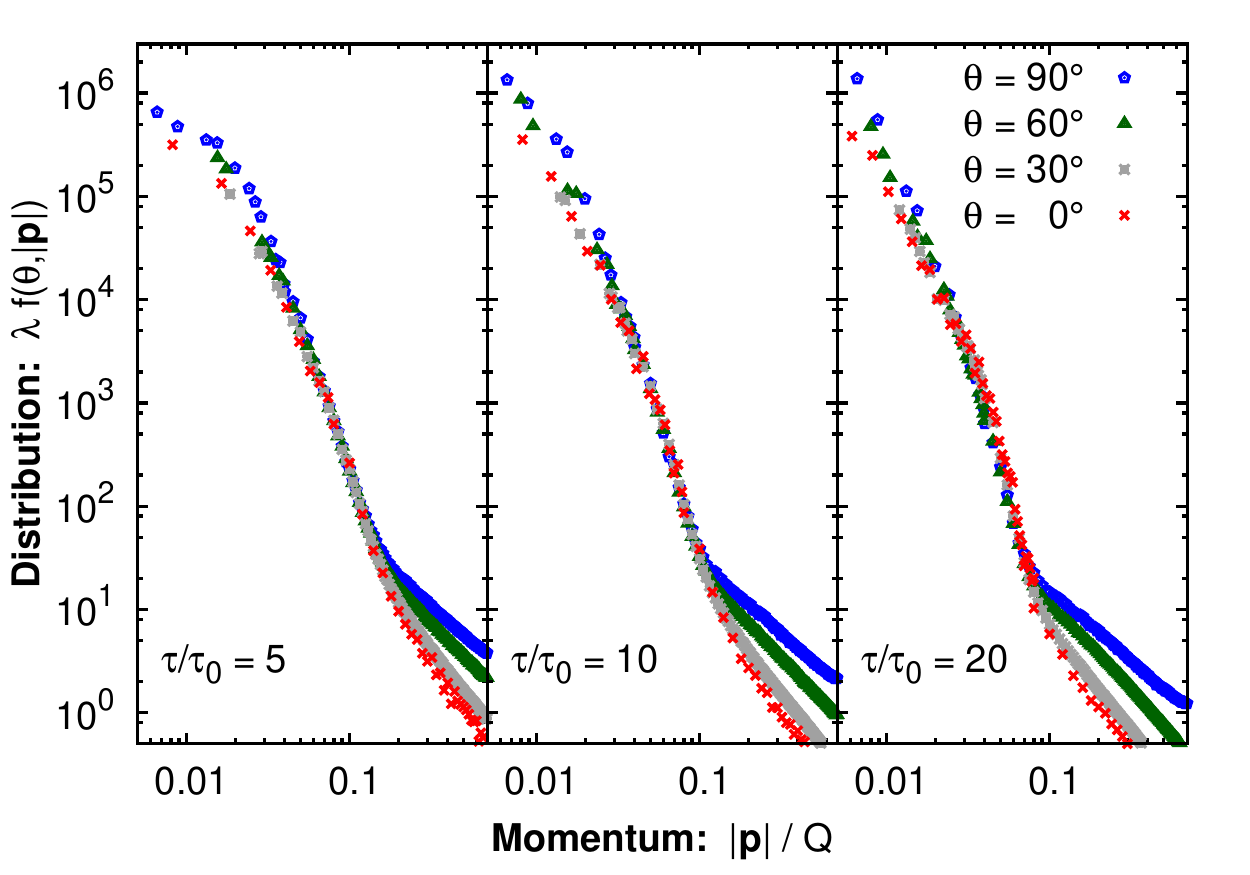}
 \caption{Distribution function for different  polar angles $\theta = \arctan (p_T/p_z)$ as a function of the momentum $|\mathbf{p}|= \sqrt{p_T^2 + p_z^2}$ at different times $\tau/\tau_0 = 5,\,10,\,20$.  While the distribution is highly anisotropic at high momenta, the infrared sector remains approximately isotropic.}
  \label{fig:isotropyPlot}
\end{figure}

To better understand the nature of the inertial range at low momenta, we analyze the dependence of the single particle distribution $f(p_T,p_z,\tau)$ on the polar angle $\theta = \arctan (p_T/p_z)$. In Fig.~\ref{fig:isotropyPlot}, we show the phase-space distribution for different values of $\theta$ as a function of the absolute momentum $|\mathbf{p}|= \sqrt{p_T^2 + p_z^2}$. While the spectrum is highly anisotropic at high momenta, one observes that the strong enhancement of the infrared region is angle independent to within the accuracy of the numerical simulations. Therefore typical transverse and longitudinal momenta evolve  isotropically, which gives 
\begin{align}
 \beta = \gamma
\end{align}
for the scaling exponents defined in Eq.~(\ref{mat:scaling}), in the IR inertial momentum region. 

When comparing the values of the mass $m(\tau)$ in Fig.~\ref{fig:dispersion-relation} with the typical momenta for the low momentum fixed point in Figs.~\ref{fig:spectrum-pz0-BoxIC} and~\ref{fig:isotropyPlot}, we find that the low momentum inertial range is always bounded by the effective mass $|\mathbf{p}| \lesssim m(\tau)$. Hence the dynamics of this region is governed by nonrelativistic physics with the time dependent dynamically generated mass $m(\tau)$. Moreover from Figs.~\ref{fig:spectrum-pz0-BoxIC} and~\ref{fig:isotropyPlot} one observes that the low momentum region entirely involves nonperturbatively large occupation numbers with at least $\lambda f \gtrsim 5$.

\begin{figure}[tp!]						
 \centering
 \includegraphics[width=0.5\textwidth]{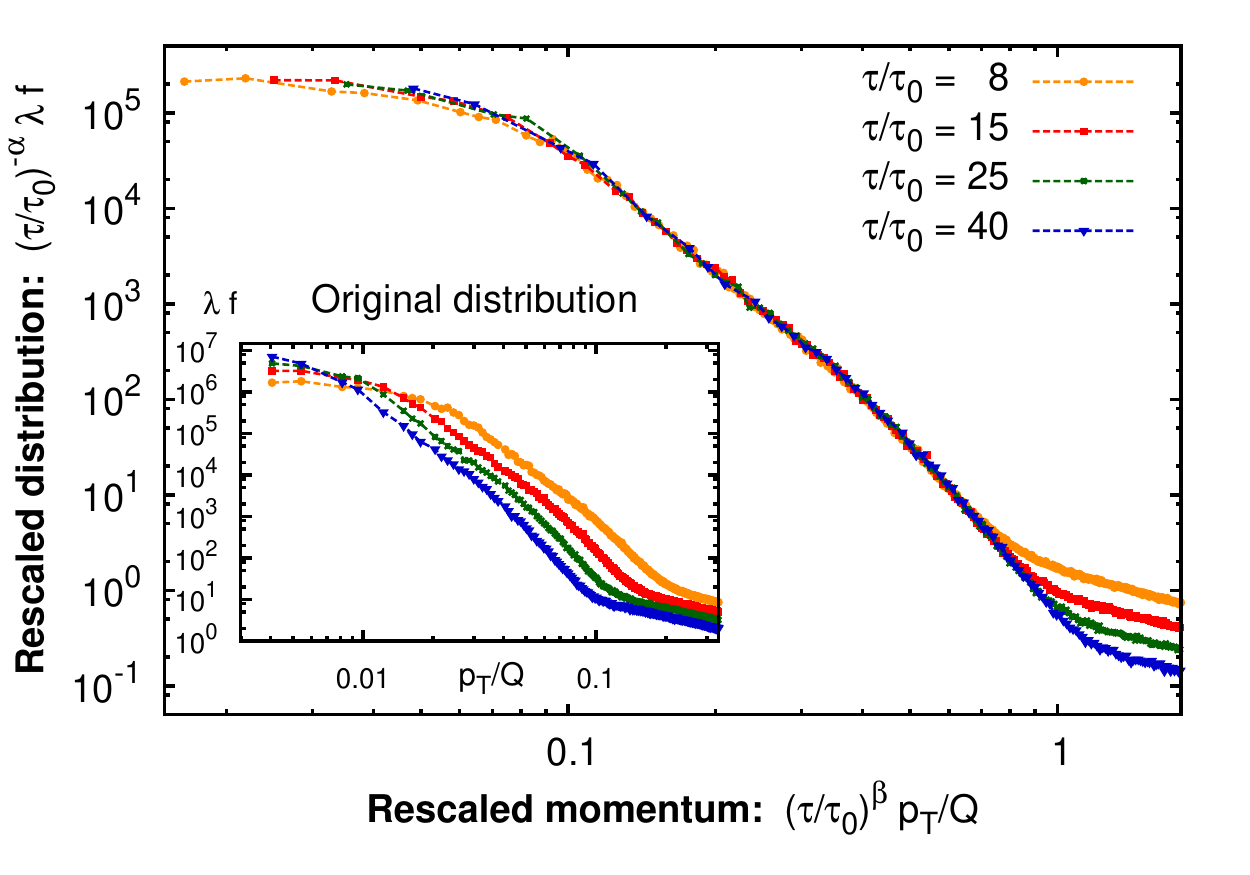}
 \caption{Rescaled distribution function of the expanding scalar system at low momenta as a function of the rescaled transverse momentum for different times. Spectra at different times collapse onto a single curve, demonstrating the self-similarity of the infrared cascade. For comparison, the original distributions without rescaling are shown in the inset.}
  \label{fig:scaling-low-momenta}
\end{figure}

\begin{figure}[tp!]						
 \centering
 \includegraphics[width=0.5\textwidth]{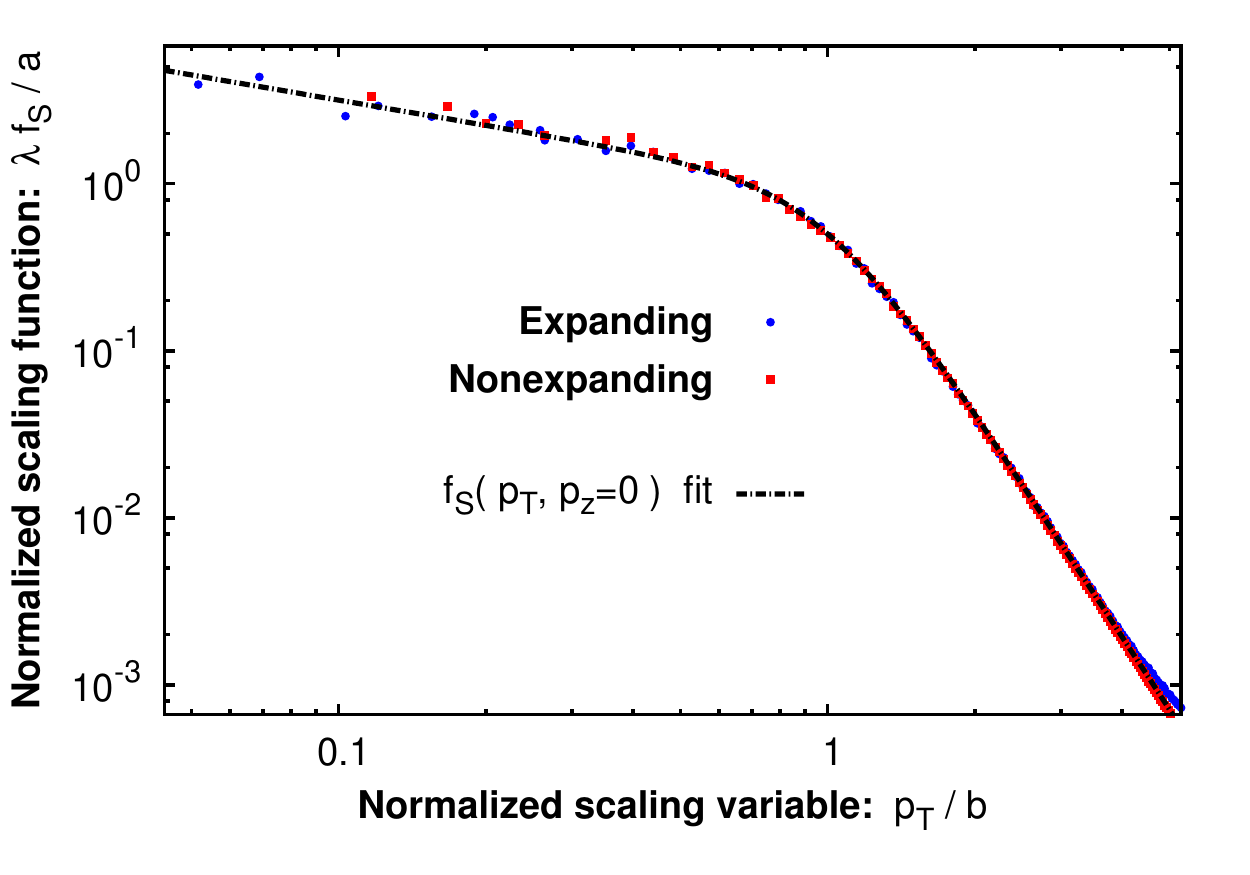}
 \caption{The normalized fixed point distribution $f(p_T,p_z=0)$ at time $2 \,\tau_0$. Also shown is the distribution function for a  two component scalar field theory in a static box, previously considered in Ref.~\cite{Orioli:2015dxa}. Both spectra can be well described by the functional form given by Eq.~(\ref{mat:fs-low-momenta}) as indicated by the black dashed line.}
  \label{fig:comparison-fs-exp-nonexp}
\end{figure}

We now turn to the universal quantities $\alpha$, $\beta$, $\gamma$ and the scaling function $f_S(p_T,p_z)$ of the low momentum scaling region. To access the scaling regime earlier, we increase the initial amplitude to $n_0 = 125$. Since we need to be sensitive to the low momentum region, we use a large lattice with $768^2 \times 512$ lattice points and lattice spacings $Q a_T = 2$ and $a_\eta = 1.5\times 10^{-4}$.

In Fig.~\ref{fig:scaling-low-momenta} we show the rescaled transverse momentum distribution $\tau^{-\alpha}f(p_T,p_z=0,\tau)$ as a function of the rescaled transverse momentum $\tau^{\beta} p_T$. The inset gives the curves at different times without rescaling for comparison. With $\alpha = 1$ and $\beta = 2/3$ the rescaled curves at different times lie remarkably well on top of each other. According to Eq.~(\ref{mat:scaling}), this reveals a self-similar evolution of the distribution function $f_S$. If particle number is conserved, one has $\tau\,n = \tau N\int d^3 p/(2\pi)^3 f = const$ which requires $\alpha - 3\beta = -1$ consistent with the values we used. The movement of the region to lower momenta, as can be seen in the inset of Fig.~\ref{fig:scaling-low-momenta}, is consistent with the positive value of the exponent $\beta > 0$  observed and can be interpreted as being due to an inverse particle cascade feeding the Bose condensate at zero momentum.

To gain insight into the statistical and systematic errors for the exponents, we follow Ref.~\cite{Orioli:2015dxa} and investigate how the values of the exponents $\alpha$ and $\beta$ depend on the reference time $\tau_{\text{ref}}$ at which we start our self-similarity analysis. To this end, we compare the rescaled distribution function at $\tau_{\text{ref}}$ with the rescaled distribution function at different times up to $\tau/\tau_{\text{ref}} \leq 5$, and extract the best values for the exponents together with an estimate of the error (see Refs.~\cite{Orioli:2015dxa,Berges:2013fga} for more details on the method). 

In Fig.~\ref{fig:scaling-low-momenta-evolution}, we show the extracted values for $\beta$ and the combination $\alpha - 3\beta$ as functions of the reference time. Also shown are the values $-1$ and $2/3$ as gray dashed lines; the former corresponds to particle number conservation and the latter to an elastic scattering process with a resummed matrix element. (This effective elastic scattering interpretation of the numerical data will be discussed shortly in Sec.~\ref{sec:infrared-explanation}.) One finds that both of these quantities approach the gray lines.  We thus have
\begin{align}
 \alpha = 1\;, \quad \beta = \gamma = \frac{2}{3}
 \label{mat:exponents-low-momenta}
\end{align}
in the vicinity of the nonthermal fixed point with an error estimate of about $10\%$ for $\beta$ and typically $10 - 15 \%$ for $\alpha - 3\beta$, as can be 
determined from the error bars in Fig.~\ref{fig:scaling-low-momenta-evolution}.

We can also specify the functional form in the scaling region as 
\begin{align}
 \lambda f_S(p_T,p_z) = \frac{a}{(|\mathbf{p}|/b)^{\kappa_<} + (|\mathbf{p}|/b)^{\kappa_>}}\,,
 \label{mat:fs-low-momenta}
\end{align}
where the non-universal constants $a$ and $b$ can be fixed by the constraints $\lambda f_S(|\mathbf{p}|=b)=a/2$ and $\lambda f_S'(|\mathbf{p}|=b) = -a(\kappa_< + \kappa_>)/(4b)$. In Fig.~\ref{fig:comparison-fs-exp-nonexp} we show the normalized distribution function $f(p_T,p_z=0)$ at $\tau = 2\,\tau_0$ as well as the corresponding distribution function for a two component scalar field theory in a static box, discussed at length in Ref.~\cite{Orioli:2015dxa}. The  scaling functions of both the static and the longitudinally expanding system are given by Eq.~(\ref{mat:fs-low-momenta}). We used the values $\kappa_< = 0.5$ and $\kappa_> = 4.5$ for the scaling function in the figure but we observe that $\kappa_>$ slowly grows with time approaching the value $\sim 5$ expected also for nonrelativistic systems~\cite{Scheppach:2009wu,Nowak:2013juc}.

Indeed Ref.~\cite{Orioli:2015dxa} showed that the IR dynamics of a scalar field theory in a static geometry is governed by effectively nonrelativistic dynamics with a nonperturbative mechanism that describes an inverse particle cascade. While the IR scaling function in Eq.~(\ref{mat:fs-low-momenta}) can be observed in both expanding and nonexpanding scalar systems, the temporal exponents $\alpha = 3/2$ and $\beta = 1/2$ for the static box in three spatial dimensions differ from the corresponding values of $\alpha=1$ and $\beta=2/3$ for the expanding theory. This difference stems from the different geometry; in the expanding case, one has a time dependent effective mass and particle number conservation gives $\tau\,n = const$ instead of $n=const$ for the fixed box. However these differences do not alter the underlying dynamics in a fundamental way.  We will show that the IR dynamics of the expanding system, like that of the static case, is consistent with a description in terms of a vertex-resummed kinetic theory.




\subsection{Low momentum region from vertex-resumed kinetic theory}
\label{sec:infrared-explanation}

We will now extract the exponents $\alpha$, $\beta$ and $\gamma$ for the inertial range at low momenta from a vertex-resumed kinetic theory and show that their values are consistent with our numerical results. We will subsequently relate these universal exponents to the dynamics that leads to the formation of the Bose-Einstein condensate.

\subsubsection{Universal exponents}

The kinetic equation for the distribution function $f$ in the longitudinally expanding background reads as 
\begin{align}
 \left( \partial_\tau - \frac{p_z}{\tau}\partial_{p_z} \right) f(p_T,p_z,\tau) = C[f](p_T,p_z,\tau)\,,
 \label{mat:kinetic-equation}
\end{align} 
with the collision integral $C[f](p_T,p_z,\tau)$ representing the scattering process and with the redshift term $-p_z\,\partial_{p_z} f/\tau$ that follows from the  boost invariance of the system in the longitudinal direction~\cite{Mueller:1999pi,Kurkela:2011ub,Berges:2013eia}. 

Since the occupation numbers are nonperturbatively large in the low momentum region $f \gtrsim 1/\lambda$, a perturbative approach is problematic. However for an $N$ component scalar theory, one can employ a $1/N$ expansion of the two particle irreducible (2PI) effective action~\cite{Berges:2004yj,Aarts:2002dj}. To NLO in this $1/N$ expansion, the temporal dynamics of the distribution function can be expressed as a kinetic equation with the vertex-resummed collision kernel introduced in Refs.~\cite{Berges:2010ez,Orioli:2015dxa}. Specifically, in this approach, all corrections to the coupling constant of the $N$-component scalar field theory are included in an effective coupling $\lambda_{\text{eff}}$. The collision integral for this effective kinetic description in the infrared describes an elastic $2 \leftrightarrow 2$ scattering process with $\lambda_{\text{eff}}[f]$, 
\begin{align}
 C[f](p_1) = \int d\Omega^{2\leftrightarrow 2}\; \frac{\lambda_{\text{eff}}^2}{6N}\; [(f_1 + f_2) f_3 f_{4} - f_{1} f_{2} (f_{3}+f_{4})]\,.
 \label{mat:collision-NLO}
\end{align}
Here we used $f_{i} \equiv f(p_{T,i},p_{z,i},\tau) \gg 1$ with $p_1 \equiv p$ and suppressed the arguments of the functionals $\int d\Omega^{2\leftrightarrow 2}[f]$ and $\lambda_{\text{eff}}^2[f]$. Their precise forms are written in Appendix~\ref{sec:collision-integral-low-momenta}.

To extract the universal scaling exponents of the low momentum region, we follow Refs.~\cite{Micha:2004bv,Orioli:2015dxa} and study the temporal scaling behavior of each side of Eq.~(\ref{mat:kinetic-equation}). Using the scaling ansatz in Eq.~(\ref{mat:scaling}), the left hand side becomes
\begin{align}
 \tau^{\alpha-1} \left( \alpha
+\beta \,\bar{p_T}\, \partial_{\bar{p_T}} + (\gamma-1)\, \bar{p_z}\, \partial_{\bar{p_z}}\,\right) f_S(\bar{p_T},\bar{p_z}) \,,
\label{mat:kinetic-lhs-scaling}
\end{align}
where we have abbreviated $\bar{p_T} \equiv \tau^{\beta} p_T$ and $\bar{p_z} \equiv \tau^{\gamma} p_z$. Similarly one can determine the scaling property of the collision integral
\begin{align}
 C[f](p_T,p_z,\tau) = \tau^{\mu} \, C[f_S](\bar{p_T},\bar{p_z})\,,
 \label{mat:collision-scaling}
\end{align}
where $C[f_S](p_T,p_z)$ is time independent. The exponent $\mu = \mu(\alpha,\beta,\gamma,\sigma)$ characterizes the scaling behavior of the collision integral, where we also included the scaling exponent $\sigma$ of the effective mass $m(\tau) \sim \tau^{-\sigma}$. Substituting Eqs.~(\ref{mat:kinetic-lhs-scaling}) and (\ref{mat:collision-scaling}) in Eq.~(\ref{mat:kinetic-equation}) leads to the time independent fixed point equation for $f_S(p_T,p_z)$
\begin{align}
& \alpha f_S(p_T,p_z)
+\beta \,p_T\, \partial_{p_T}\, f_S(p_T,p_z) \nonumber\\
& +  \left(\gamma-1\right) p_z\, \partial_{p_z}\, f_S(p_T,p_z)
= C[f_S](p_T,p_z)\,,
\label{mat:kinetic-stationary}
\end{align}
where we have simply relabeled $\bar{p_T} \mapsto p_T$ and $\bar{p_z} \mapsto p_z$, and to the scaling relation for the temporal exponents
\begin{align}
 \alpha - 1 = \mu\,.
 \label{mat:relation-mu}
\end{align}
A scaling solution of the kinetic equation has to satisfy both equations simultaneously. A detailed discussion of the fixed point equation~(\ref{mat:kinetic-stationary}) and its solution is beyond the scope of this paper. We will focus on the scaling relation Eq.~(\ref{mat:relation-mu}) since we are primarily interested in the scaling exponents.

To determine the scaling exponent of the collision integral $\mu$, we first observe that since the inertial range at low momenta is bounded by the effective mass $m(\tau)$, the dispersion relation becomes
\begin{align}
 \omega(p_T,p_z,\tau) \simeq m(\tau) + \frac{|\mathbf{p}|^2}{2m(\tau)}\,.
 \label{mat:omega-large-mass}
\end{align}
Because the leading mass term cancels in intermediate steps, one has to determine the scaling property of the momentum $|\mathbf{p}|^2$. As our numerical results in Fig.~(\ref{fig:isotropyPlot}) show, the momentum distribution in the IR is isotropic. Hence we can assume that longitudinal and transverse momenta as well as the modulus all scale equally with $\beta = \gamma$.

The precise form of the exponent $\mu$ is derived in Appendix~\ref{sec:collision-integral-low-momenta} and reads as 
\begin{align}
 \mu = \alpha - 2\beta + \sigma\,,
 \label{mat:mu-exponent}
\end{align}
where the reader will recall that the scaling exponent $\sigma$ controls the temporal evolution of the effective mass. From the scaling relation in Eq.~(\ref{mat:relation-mu}), and using the isotropy condition, one obtains 
\begin{align}
 \beta = \gamma = \frac{1+\sigma}{2}\,.
 \label{mat:beta-prime-exponent}
\end{align}
Since our numerical results show that the mass scales with $\sigma = 1/3$, the scaling relation deduced from the effective kinetic theory results in momentum exponents that decrease with $\beta = \gamma = 2/3$, which is consistent with the numerical result we obtained from classical-statistical simulations. 

To determine the scaling exponent $\alpha$ from the vertex-resummed kinetic theory, we need a second condition. The form of  the elastic scattering kernel in Eq.~(\ref{mat:kinetic-equation}) suggests that particle number (defined as the integral of $f$ over phase space) is conserved.  One can confirm that the collision integral (\ref{mat:collision-NLO}) vanishes when we integrate\footnote{Recall that $\nu = \tau\, p_z$.} $N\int d^{2} p_T\,d\nu/(2\pi)^3 C[f] = 0$. Further, the corresponding integral over the left hand side of Eq.~(\ref{mat:kinetic-equation}) can be identified with the total time derivative of the distribution function
\begin{align}
 \left( \partial_\tau - \frac{p_z}{\tau}\partial_{p_z} \right) f(p_T,p_z,\tau) = \frac{d}{d\tau} f(p_T,\nu/\tau,\tau)\,,
\end{align}
and therefore integrating the kinetic equation (\ref{mat:kinetic-equation}) leads to
\begin{align}
 0 = \frac{d}{d\tau}~ N\int \frac{d^{2} p_T\;d\nu}{(2\pi)^3} f(p_T,\nu/\tau,\tau) = \frac{d}{d\tau}~\tau\, n\,.
\end{align}
Thus $\tau\, n = const$ in the IR inertial range;  from the self-similar form of Eq.~(\ref{mat:scaling}), one then obtains 
\begin{align}
 \alpha = 2\beta + \gamma -1\,.
\end{align}
With Eq.~(\ref{mat:beta-prime-exponent}), this leads to 
\begin{align}
 \alpha = (1+3\,\sigma)/2\,.
\end{align}
Again, plugging in $\sigma = 1/3$ we obtain $\alpha = 1$, which is again consistent with the result obtained from the classical-statistical numerical simulations. 
One may therefore conclude that the vertex-resummed kinetic theory obtained from the NLO $1/N$ expansion in the IR inertial range is consistent with first principles classical-statistical numerical simulations of the expanding scalar theory.

\subsubsection{Condensate formation}

With the scaling properties of the IR inertial region, we are now able to further analyze the power law growth of the zero mode $F_0(\tau) \equiv F(p_T=0,p_z=0,\tau)$ in Fig.~\ref{fig:condensation} for sufficiently large volumes. The correlation function $F$ can be decomposed into a particle and condensate part according to 
\begin{align}
 &F(p_T,\nu,\tau) = \frac{f(p_T,p_z,\tau)}{\tau\;\omega(p_T,p_z,\tau)} + (2\pi)^3 \delta^{(2)}(\mathbf{p_T})\delta(\nu)\phi_0^2(\tau)\,.
 \label{mat:F-f-omega-phi-relation}
\end{align}
For finite momenta, the ``particle" part (c.f. App.~\ref{sec:asymptotics}) of this expression provides an alternative definition of the distribution function. Our expression in Eq.~(\ref{mat:F-f-omega-phi-relation}) is essentially the same as the one derived previously for a static box~\cite{Orioli:2015dxa}, except for the additional factor $1/\tau$ due to the expanding metric. We have checked numerically that it converges to the previous definition in Eq.~(\ref{mat:distr-func-definition}) with increasing accuracy in time. 

When considering overoccupation initial conditions (\ref{mat:fluct-IC}) there is no condensate initially but the inverse particle cascade populates the zero mode  according to the relation
\begin{align}
\tau^{2/3} F_0 \sim \frac{f(p=0,\tau)}{\omega(p=0)\,\tau^{1/3}} \sim \tau^{\alpha + \sigma -1/3} \sim \tau^{\alpha}\,,
\label{eq:QWDF}
\end{align}
where we have used the scaling exponent of the mass $\sigma = 1/3$. Denoting the value of the rescaled zero momentum correlator $(\tau/\tau_i)^{2/3} F_0(\tau)/V$ at the initial time $\tau_i$ of the self-similar regime as $f(p=0,\tau_i)/(\tau_i m(\tau_i)\,V)$ and its final value at the condensation time $\tau_c$ as $\tilde{\phi}_0^2(\tau_c)= (\tau_c/\tau_i)^{2/3} \phi_0^2(\tau_c)$) the condensation time $\tau_c$ then follows from the power law dependence of the zero mode in Eq.~\ref{eq:QWDF} as 
\begin{equation}
\tau_c \,\simeq\,  \tau_i \, \left(\frac{\tilde{\phi}_0^2(\tau_c)\,\tau_i\,m(\tau_i)}{f(p=0,\tau_i)}\right)^{1/\alpha} \, V^{1/\alpha} \, .
\label{eq:condens-time-estimate}
\end{equation}
Hence, we have $\tau_c \sim V^{1/\alpha}$, which is the same relation as for the non-expanding theory~\cite{Orioli:2015dxa}. Most importantly it diverges with increasing volume as $\tau_c \sim V$ for the expanding theory close to its attractor.

We can also estimate the number of particles in the condensate $n_c$ beyond the onset of condensation. By use of the decomposition in  Eq.~(\ref{mat:F-f-omega-phi-relation}) one finds
\begin{align}
 n_c \sim \omega(p=0)\,\phi_0^2 \sim \tau^{-1}\,,
\end{align}
where in the last step we used the scaling properties of the mass and the condensate. Hence once a condensate is created, its particle number is approximately conserved and its evolution effectively decouples from the rest of the system. This is certainly not true for times prior to the formation of the condensate, where the low momentum region plays an essential role.




\begin{figure}[tp]
\begin{minipage}[left]{0.5\textwidth}
 \includegraphics[width=\textwidth]{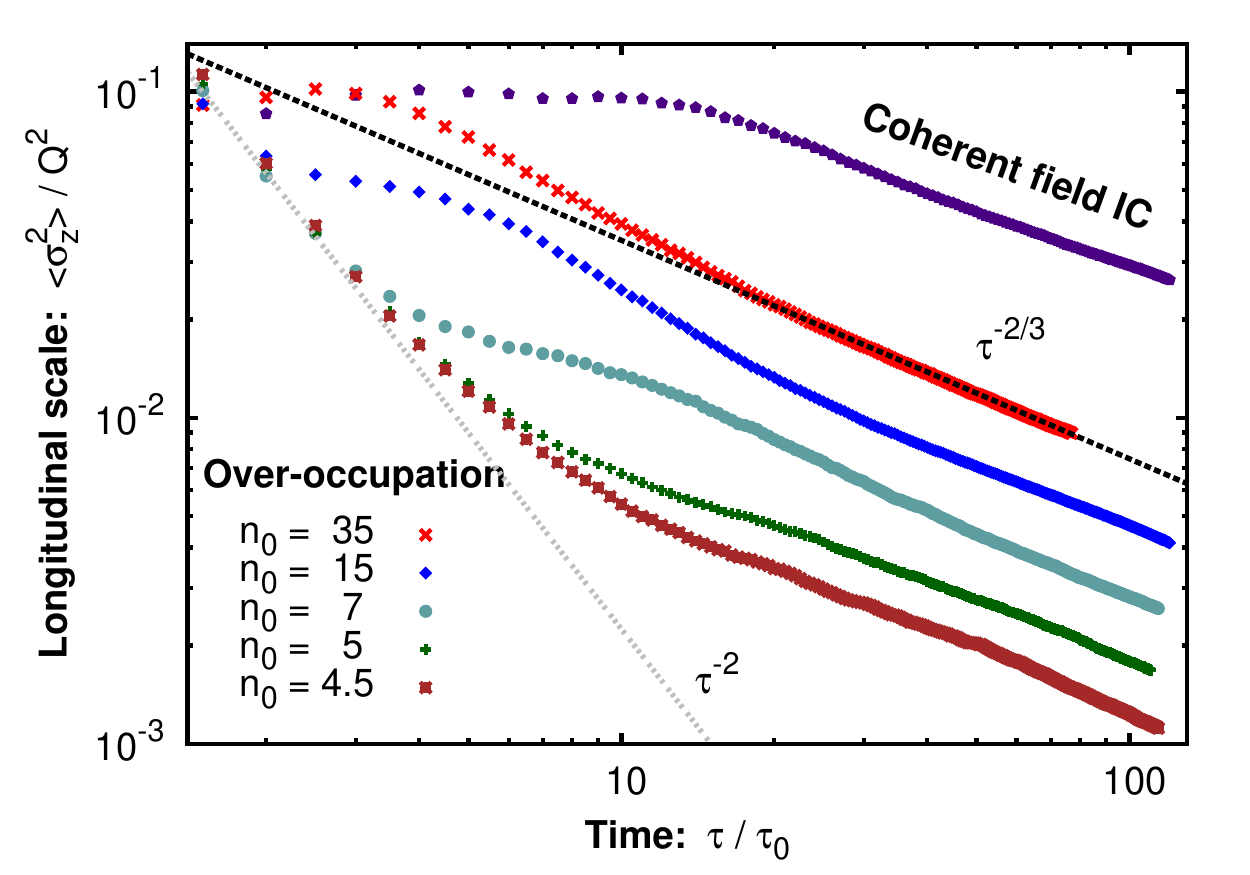}
\end{minipage}
\begin{minipage}[left]{0.5\textwidth}
 \includegraphics[width=\textwidth]{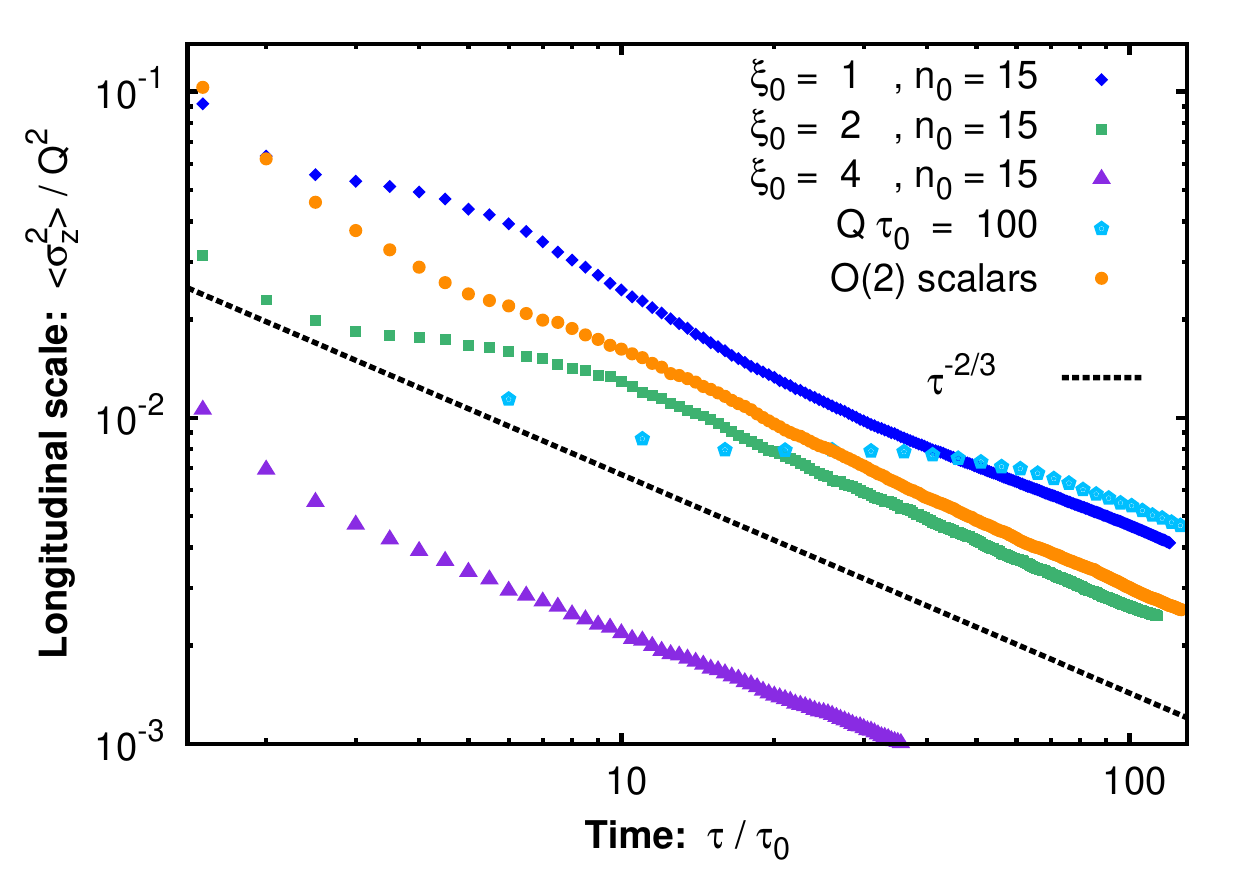}
\end{minipage}
 \caption{Time evolution of the longitudinal momentum scale $\langle\sigma_z^2\rangle (\tau)$ for different initial conditions. Top figure: After a short transient behavior, the dynamics becomes insensitive to different values of $n_0$ and all curves follow an approximate $\tau^{-2/3}$ power law dependence indicated by the black dashed line. Comparison to the gray dashed free streaming ($\sim \tau^{-2}$) curve shows the significance of momentum broadening. 
 Bottom figure: Three of the curves show the insensitivity of the late time behavior to variations of the initial momentum anisotropy $\xi_0$. The curve labeled $Q\tau_0=100$ corresponds to a different initialization time and demonstrates insensitivity of late time results to the starting time. The plot labeled $O(2)$ scalars shows the results for two component scalars. Since all the other curves are for $N=4$ scalars, this curve demonstrates that late time results are independent of the number of degrees of freedom.}
\label{fig:sigmaz-compare-IC}
\end{figure}

\subsection{Intermediate momenta}
\label{sec:intermediate-momenta}

We now turn to a more detailed analysis of the intermediate momentum sector characterized by the $\sim 1/p_T$ power law in  Fig.~\ref{fig:spectrum-pz0-BoxIC}. In contrast to the soft sector, the spectrum at intermediate momenta is anisotropic; it is more efficiently described by considering the dependence on the longitudinal and transverse momenta separately.

Regarding the transverse momentum dependence, we discussed in \cite{Berges:2014bba} that in this case, there is effectively neither a particle nor an energy flux in transverse momentum. Instead the energy and particle number per transverse momentum mode $\tau\,dn/dp_T$ and $\tau\,d\varepsilon/dp_T$ are time independent and thus conserved at each transverse momentum separately. Additionally, $\tau\,dn/dp_T$ is uniformly distributed over transverse momenta, 
\begin{align}
 \tau\,\lambda\frac{dn}{dp_T} \sim \tau\;p_T\int\frac{d p_z}{2\pi}\,\lambda f(p_T,p_z,\tau) = const\,,
 \label{mat:reduced-distribution}
\end{align}
up to a rapid fall-off for $p_T \gtrsim 2\,Q$. 

The local conservation of both energy and particle number is a general property of the anisotropic scaling regions of the longitudinally expanding scalar field theory and has also been observed in Ref.~\cite{Berges:2013eia} for the non-Abelian gauge theory. In contrast, there is no single scaling solution conserving both energy and particle number for isotropic systems~\cite{Micha:2004bv}. Instead only one conservation law constrains the scaling solution. This has been observed not only for the scalar theory in a static geometry~\cite{Micha:2004bv,Berges:2008wm,Nowak:2010tm} but also within the low momentum inertial range of the expanding scalar theory where local particle number conservation leads to an inverse particle cascade.

In the language of the self-similar evolution in Eq.~(\ref{mat:scaling}), the local conservation of energy and particle number in the anisotropic scaling regime gives rise to the scaling relations
\begin{align}
 \alpha - \gamma + 1 = 0 \,, \qquad \beta = 0\,.
 \label{mat:relations-a-b-g}
\end{align}
With these constraints taken into account, the relevant dynamics in this regime is that of longitudinal momentum broadening.

Momentum broadening in the longitudinal direction can be efficiently analyzed in terms of the characteristic longitudinal momentum scale
\begin{align}
 \sigma_z^2(p_T,\tau) = \frac{\int dp_z ~p_z^2 f(p_T,p_z,\tau)}{\int dp_z f(p_T,p_z,\tau)}\,.
 \label{mat:sigmaz}
\end{align}
In Fig.~\ref{fig:sigmaz-compare-IC} we show the time evolution of $\langle\sigma_{z}^2\rangle(\tau)$ (that is $\sigma_z^2(p_T,\tau)$ averaged over transverse momenta within the intermediate inertial range ii) for various simulation parameters. The top panel shows the behavior under variations of the initial overoccupancy $n_{0}$, while the bottom panel corresponds to variations of the initial anisotropy $\xi_{0}$, the initial time $Q\, \tau_{0}$ as well as the number of field components. While for different initial conditions the early time dynamics ranges from free streaming to an approximately constant behavior, one clearly observes that the late time behavior becomes insensitive to all of these aspects. Ultimately all curves follow a universal scaling  behavior in time. 

According to Eq.~(\ref{mat:scaling}) the scaling behavior of  $\sigma_{z}$ is characterized in terms of the scaling exponent $\gamma$, 
\begin{align}
\sigma_{z}^2(p_T,\tau)\sim \tau^{-2\gamma}\, .
\end{align}
We have checked that this scaling holds for all transverse momenta within the inertial range. By performing a combined analysis of all data sets, we obtain an estimate of the scaling exponent
\begin{align}
 2 \gamma = 0.66 \pm 0.05\;.
\end{align}
A comparison with the data in Fig.~\ref{fig:sigmaz-compare-IC}, shows that the observed scaling behavior is very well reproduced by an approximate $\tau^{-2/3}$ power law dependence.

\begin{figure}[tp!]						
 \centering
 \includegraphics[width=0.5\textwidth]{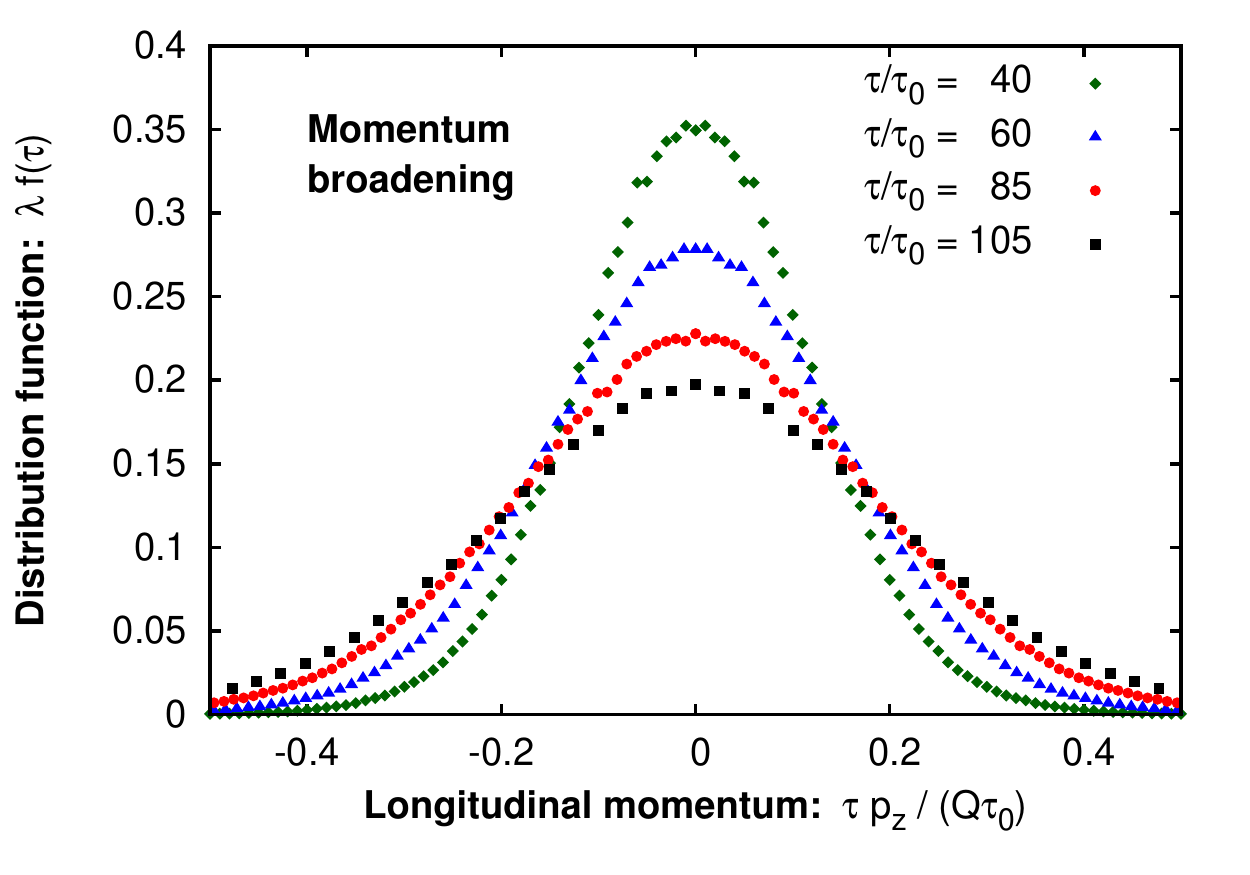}
   \caption{\label{fig:mom-broadening} Dependence of the single particle distribution function on the longitudinal momentum variable $\nu = \tau p_z$ at $p_T = Q/2$ for different times of the evolution. Momentum broadening leads to a redistribution of the spectrum from lower to higher $\nu$.}
\end{figure}   

\begin{figure}[tp!]
\centering   
 \includegraphics[width=0.5\textwidth]{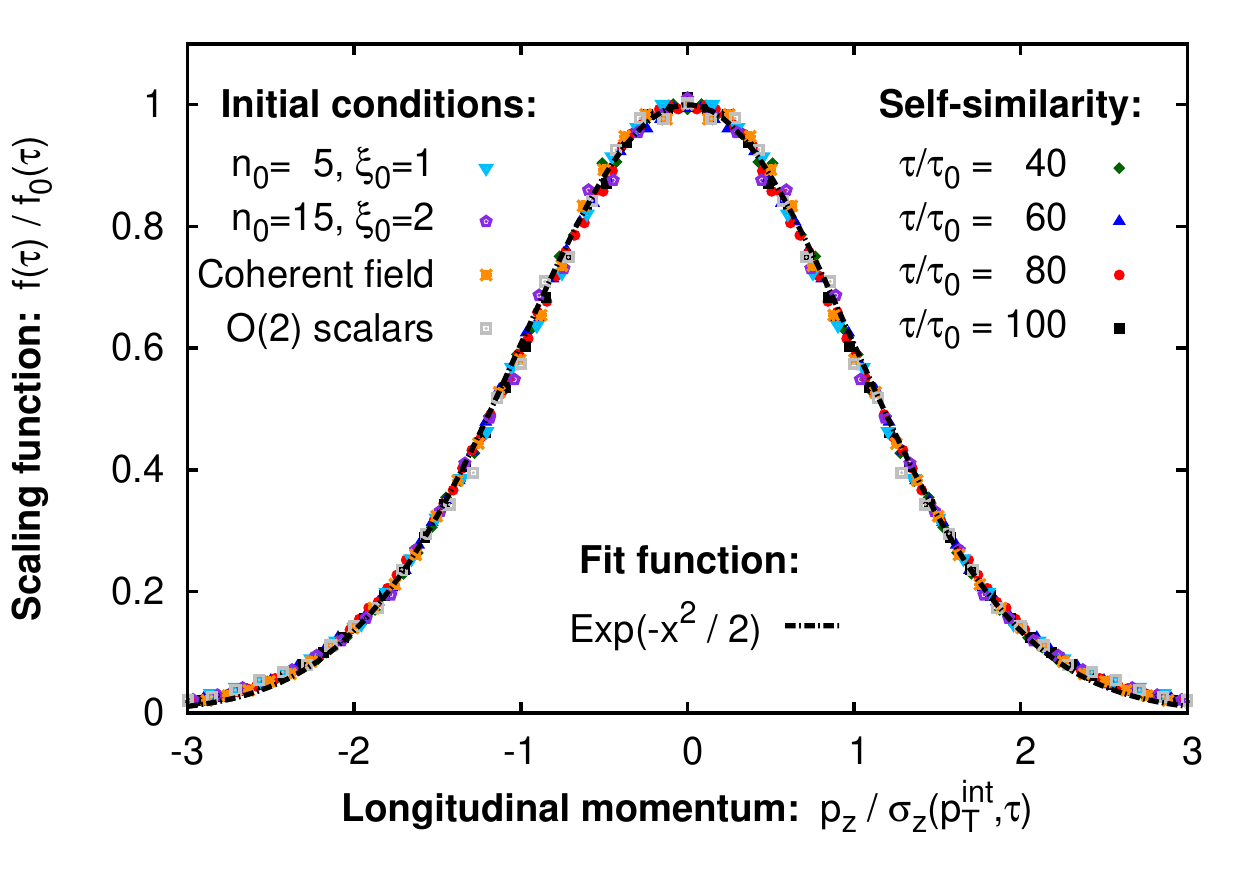}
  \caption{\label{fig:self-similar-gauss}Normalized distribution $f_{S}(p_z/\sigma_z)$ for the fixed point at typical \emph{intermediate} transverse momenta $p_T^{{\rm int.}}$. The rescaled spectra of Fig.~\ref{fig:mom-broadening} for $n_0 = 35$ and $\xi_0 = 1$ at different times collapse onto a single curve, proving the self-similarity of the evolution (right labels). As shown, the functional form of the fixed point distribution is independent of the initial conditions (left labels) and can be described by a normalized Gaussian distribution (black dashed curve).}
\end{figure}

We can also analyze the dynamics of momentum broadening on a more microscopic level by directly considering the single particle distribution at a typical transverse momentum $p_T^{\rm{int.}}$ within the intermediate inertial range ii). Our results for the time evolution of the longitudinal spectrum are presented in Fig.~\ref{fig:mom-broadening}, which shows snapshots of the distribution at $p_T^{{\rm int.}} = Q/2$ for different times. The spectrum is shown as a function of the rapidity wave number $\nu=\tau p_z$ on the horizontal axis, which effectively amounts to undoing the redshift of longitudinal momenta. While for a noninteracting system the distribution becomes time independent when plotted as a function of $\nu$, this is clearly not the case for the longitudinally expanding scalar theory. Instead, Fig.~\ref{fig:mom-broadening} shows a clear broadening of the longitudinal momentum distribution as a function of time along with a decreasing amplitude $f_{0}(\tau)=f(p_T = p_T^{{\rm int.}},p_z=0,\tau)$.

Once we take into account the decrease of the overall amplitude of the distribution according to
\begin{align}
f_{0}(\tau) \sim \tau^{\alpha}
\end{align}
with $\alpha\simeq-2/3$ determined by the scaling relation in Eq.~(\ref{mat:relations-a-b-g}) and normalize the longitudinal momenta by the typical momentum scale $p_z/\sigma_{z}(p_T^{{\rm int.}},\tau)$, the rescaled distribution approaches a stationary (time independent) fixed point $f_{S}(p_z/\sigma_{z})$. This is shown in Fig.~\ref{fig:self-similar-gauss}, where we present the rescaled distribution $f(p_T=p_T^{{\rm int.}},p_z,\tau)/f_{0}(\tau)$ as a function of the rescaled longitudinal momentum $p_z/\sigma_{z}(p_T^{{\rm int.}},\tau)$. The fact that the results at different times are indistuinguishable confirms that the dynamics of momentum broadening can be accurately described by a self-similar scaling behavior as in Eq.~(\ref{mat:scaling}).

The functional form of the fixed point distribution is well described by a Gaussian distribution of width $\sigma_{z}$, which is shown as a black dashed line. One can also verify explicitly from Fig.~\ref{fig:self-similar-gauss} that the functional form of the fixed point distribution is independent of the initial conditions as well as the number of field components. We conclude that the universal properties of the intermediate momentum regime can be summarized in terms of the scaling exponents
\begin{align}
\alpha \simeq-2/3, \quad \beta\simeq0\;,\quad  \gamma \simeq1/3\,
\end{align}
and the scaling function 
\begin{align}
 \lambda f_{S}(p_T,p_z)\simeq \frac{A}{p_T} \exp\left( -\frac{p_z^2}{2 \sigma_{z}^2} \right)\,,
\label{mat:fs-inter-momenta}
\end{align}
with a nonuniversal constant $A$. 

Remarkably, these features of the intermediate momentum regime of the expanding scalar theory are identical to those that we demonstrated previously for 
expanding overoccupied non-Abelian plasmas~\cite{Berges:2013eia,Berges:2013fga}. In \cite{Berges:2014bba}, we argued that this agreement provides strong proof of universality, since the underlying elementary microscopic dynamics of the two theories are very different. Our results here significantly strengthen the arguments presented in \cite{Berges:2014bba}.




\subsection{Hard momentum region}
\label{sec:hard-momenta}

As noted previously, the inertial range of momenta for the intermediate momentum regime shifts towards lower momenta. At late times, an additional scaling regime emerges for hard transverse momentum modes, where the $p_T$ spectrum in Fig.~\ref{fig:spectrum-pz0-BoxIC} becomes approximately independent of the transverse momentum. Since the spectrum in this regime is highly anisotropic (with the typical longitudinal momenta much smaller than the typical transverse momenta), we will again analyze the transverse and longitudinal momentum dependence separately. Just as in the case of the intermediate momentum range, we find that there is effectively neither a particle nor an energy flux in the transverse direction. Thus as was the case in our prior discussion for the intermediate momentum range, the relevant dynamics is that of longitudinal momentum broadening and the scaling relations in Eq.~(\ref{mat:relations-a-b-g}) also hold for the hard momentum sector.

The momentum broadening in this novel hard sector can be efficiently described in terms of the hard longitudinal momentum scale
\begin{align}
 \Lambda_L^2(\tau) = \frac{\left\langle \,\left( \partial_{\eta} \partial_\mu \varphi_a(\tau,\mathbf{x}_T,\eta)\right) \left(\partial_{\eta} \partial^\mu \varphi_a(\tau,\mathbf{x}_T,\eta)\right)\, \right\rangle}{\tau^2 \left\langle \,\left( \partial_\mu \varphi_a(\tau,\mathbf{x}_T,\eta)\right) \left( \partial^\mu \varphi_a(\tau,\mathbf{x}_T,\eta)\right) \,\right\rangle}\,,
\end{align}
with summation over spatial indices $\mu = x^1, x^2, \eta$ implied and where $\langle ~\rangle$ includes a volume average. 

Within the quasiparticle picture (see Eq.~(\ref{mat:F-f-omega-phi-relation})) and using $\omega(p_T,p_z,\tau) \simeq |\boldsymbol{p}|$ for hard momenta, the hard scale $\Lambda_{L}^{2}$ can be expressed as
\begin{equation}
 \Lambda_{L}^2(\tau) \simeq \frac{ \int d^2p_T \int dp_z ~p^2_{z}\; \omega(p_T,p_z,\tau)\,f(p_T,p_z,\tau)}{\int d^2p_T \int dp_z~\omega(p_T,p_z,\tau)\,f(p_T,p_z,\tau)} \,,
 \label{mat:hard-scales-pert-scalar}
\end{equation}
which is closely related to a transverse momentum integral of $\sigma_{z}^{2}(p_T)$. However a crucial difference is the additional weighting by the mode energy $\omega$, which ensures that $\Lambda_{L}^2$ is dominated by high momentum modes.

\begin{figure}[tp!]						
 \centering
  \includegraphics[width=0.5\textwidth]{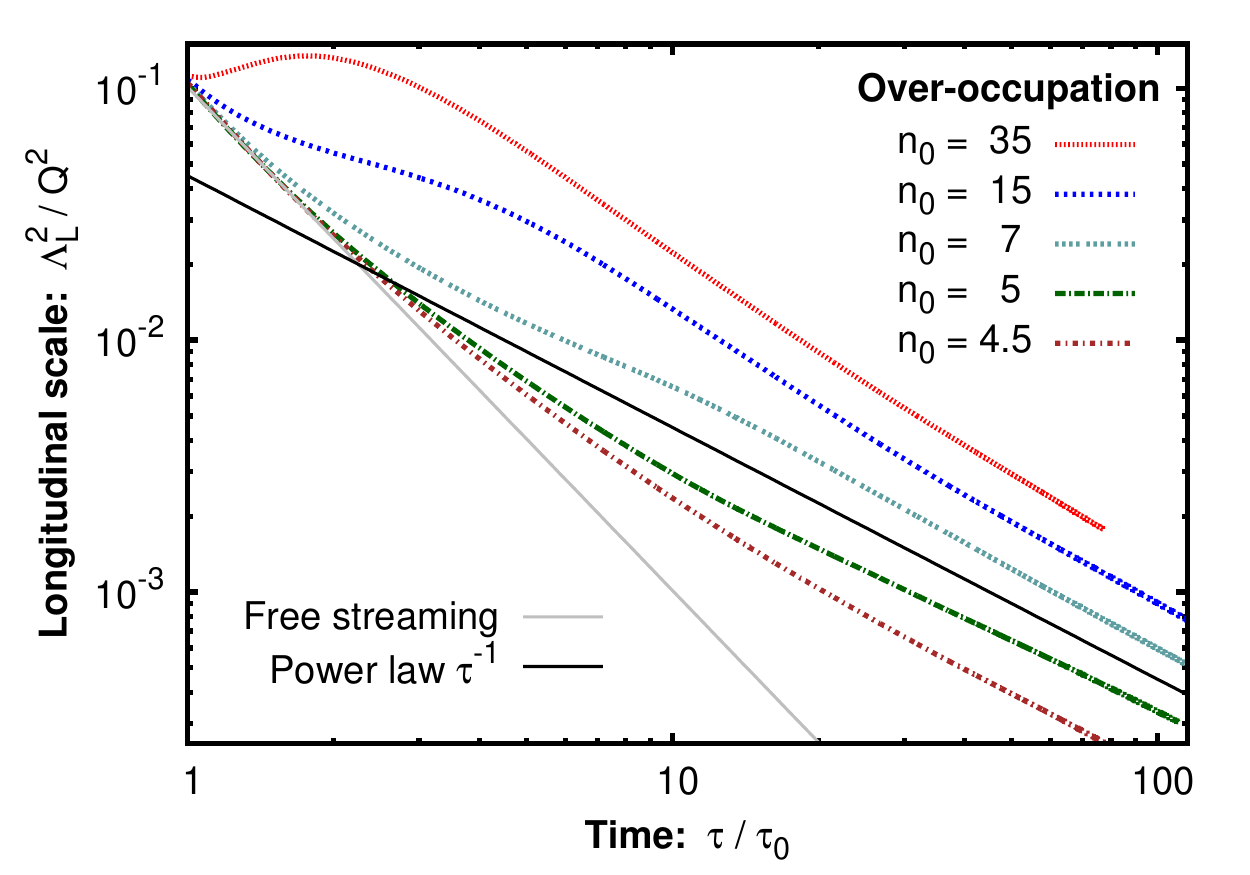}
 \includegraphics[width=0.5\textwidth]{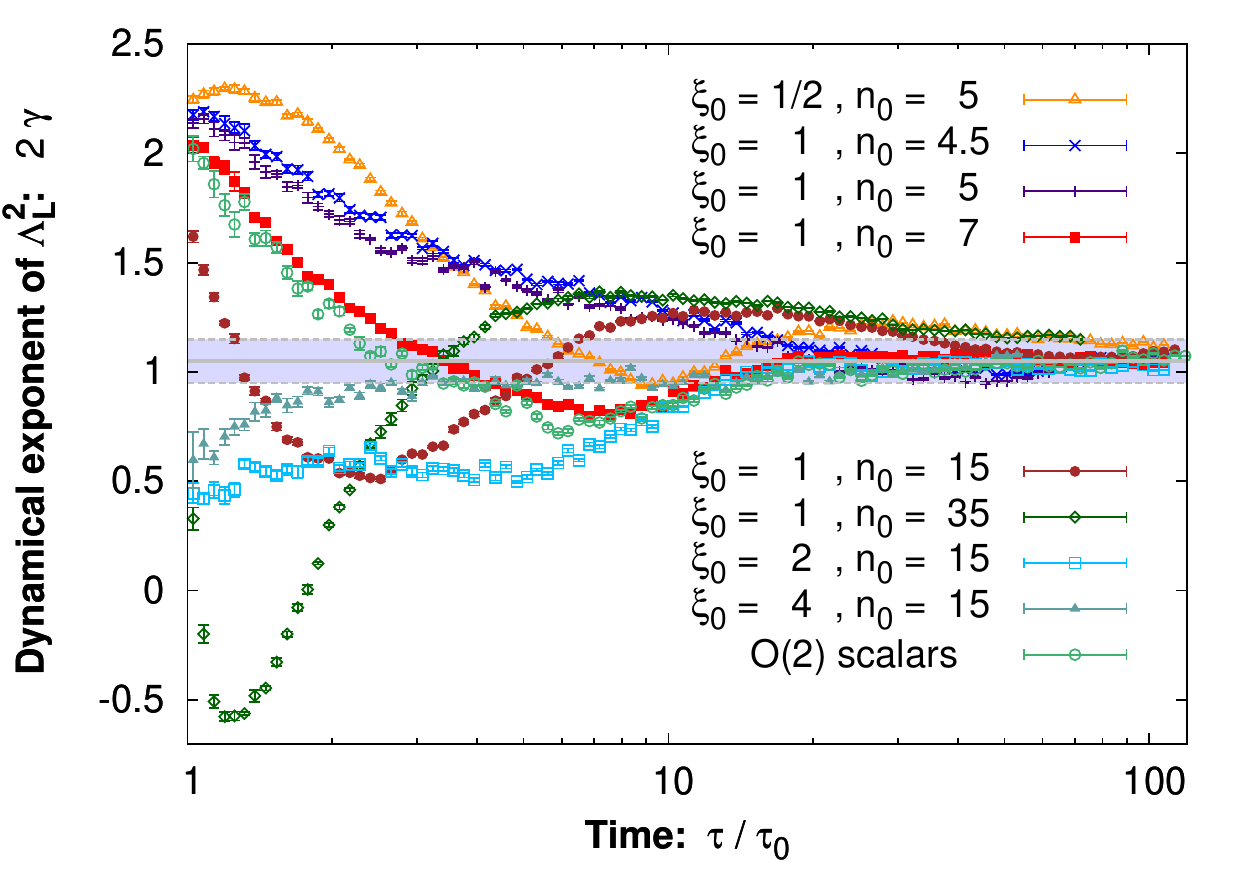}
  \caption{Upper panel: Time evolution of the longitudinal momentum scale at \emph{hard} transverse momenta for different initial conditions. At late times all curves approach a common scaling behavior (black line). 
  Lower panel: The scaling exponent $2 \gamma$ is extracted from the logarithmic derivative in the lower panel. The gray lines indicate the value $2 \gamma = 1.05 \pm 0.1$ obtained from an average over all shown simulations.}
\label{fig:HardLogScale-NXi}
\end{figure}

The time evolution of the longitudinal hard scale is shown in the upper plot of Fig.~\ref{fig:HardLogScale-NXi}. One observes that after an initial transient behavior the results for different initial conditions approach a common power law decay. By use of Eqns.~(\ref{mat:scaling}) and (\ref{mat:relations-a-b-g}), this scaling behavior is described by $\Lambda_{L}^{2}\sim \tau^{-2\gamma}$ and the dynamical exponent $\gamma$ can be extracted from the hard scale by taking the logarithmic derivative
\begin{align}
 2 \gamma(\tau) = -\frac{d\log \Lambda_L^2}{d\log \tau}\;.
\end{align}
The resulting exponent $\gamma(\tau)$ is shown in the lower plot of Fig.~\ref{fig:HardLogScale-NXi} as a function of time for different initial conditions. One observes that all curves approach a constant value,
\begin{align}
 2 \gamma = 1.05 \pm 0.1\;,
\end{align}
independent of the initial condition. By use of the scaling relations (\ref{mat:relations-a-b-g}) we infer from this analysis the approximate values of the scaling exponents
\begin{align}
\alpha \simeq -1/2\;, \qquad \beta=0\;, \qquad \gamma\simeq 1/2\;.
\end{align}
We note that the larger value of $\gamma$ compared to the intermediate momentum sector characterizes the fact that the momentum broadening is less efficient for high momentum modes. Nevertheless the observed behavior is distinctly different from free streaming and significant momentum broadening occurs throughout the entire time. Moreover we can understand the motion of the boundary between the intermediate and hard sectors (dashed purple line in Fig.~\ref{fig:spectrum-pz0-BoxIC}) by comparing the decrease of the amplitude $f_0(p_T,\tau) \equiv f(p_T,p_z=0,\tau)$ in both regions. The larger value of $\alpha$ in the hard region relative to the intermediate one implies a slower decrease of its amplitude. Therefore the boundary between both sectors moves to lower transverse momenta and increases the inertial range iii) of hard transverse momenta.

\begin{figure}[tp!]						
 \centering
 \includegraphics[width=0.5\textwidth]{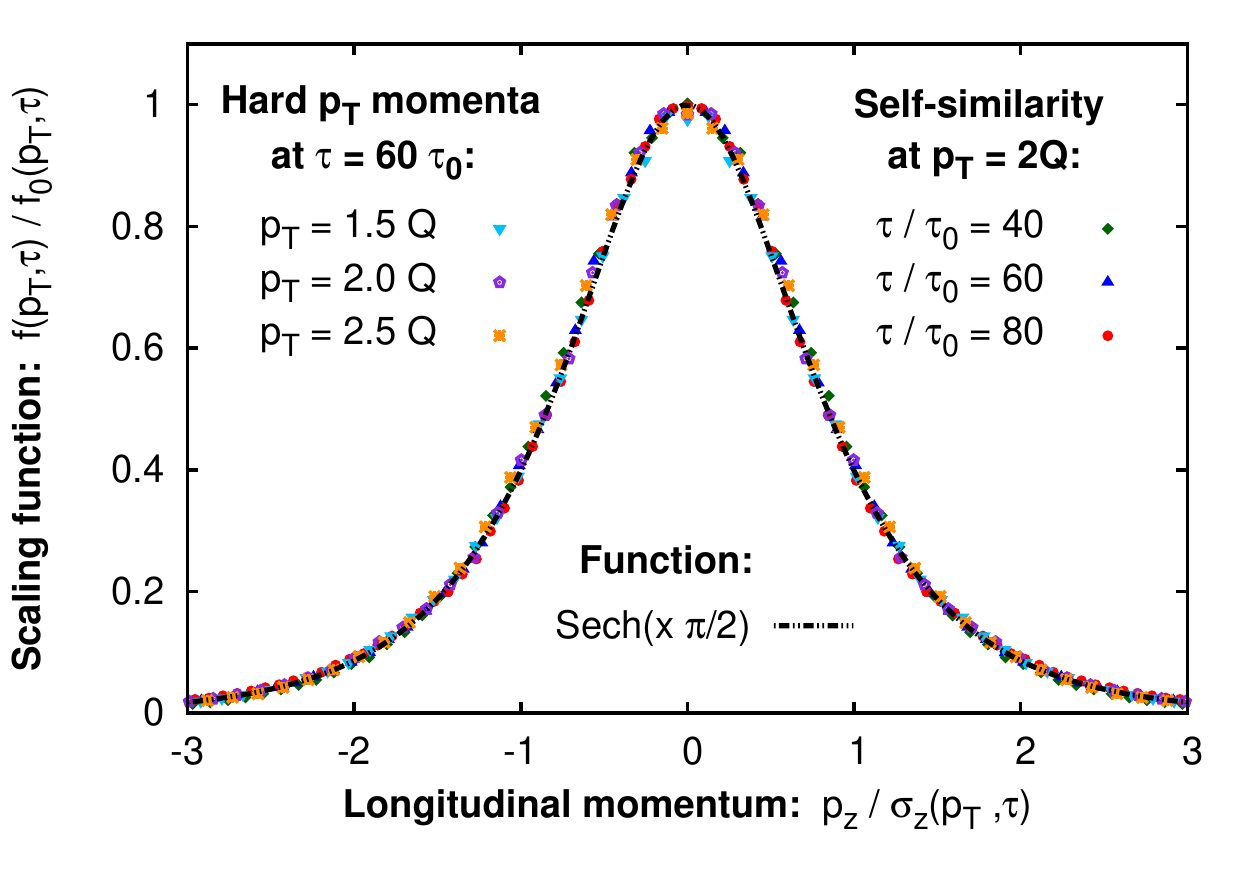}
  \caption{ \label{fig:longSpecHard} Normalized fixed point distribution $f_{S}(p_z/\sigma_z)$ for the fixed point at \emph{hard} transverse momenta. The rescaled spectra at different times collapse onto a single curve, proving self-similarity of the evolution (right labels). The functional form of the fixed point distribution is independent of the initial conditions (left labels) and can be described by a normalized hyperbolic secant function $\text{sech}(x \pi/2) = 1/\text{cosh}(x \pi/2)$.}
\end{figure}

We have also extracted the scaling function $f_S(p_z/\sigma_{z})$ following the same procedure outlined in the previous section. Our results are summarized in Fig.~\ref{fig:longSpecHard}, where we show the normalized distribution function $f(p_T,p_z,\tau)/f_0(p_T,\tau)$ as a function of the normalized longitudinal momentum $p_z/\sigma_z(p_T,\tau)$. Our results at different times agree with each other and demonstrate the emergence of self-similarity with a time independent scaling function. We have also included the results for different hard transverse momenta $p_T > Q$, which show that the scaling function has the same form over the entire inertial range of momenta.

One observes from Fig.~\ref{fig:longSpecHard} that the functional form of the scaling function is quite different from the Gaussian shape observed for the intermediate momentum fixed point. Compared to a Gaussian, the distribution in Fig.~\ref{fig:longSpecHard} features a smaller width and larger tails. This functional form is well described by a hyperbolic secant $\text{sech}(x\;\pi/2)$, which is also shown in the figure.




\section{Implications of self-similarity for bulk anisotropy}
\label{sec:implications}

Since the scalar field theory shows a rich dynamical structure, with different scaling behaviors observed in different momentum regimes, it is interesting to investigate how the interplay of the different sectors affects the evolution of bulk observables. We will focus on the bulk anisotropy of the system and investigate the time evolution of the ratio of the longitudinal and transverse pressures. 

The energy momentum tensor of the scalar field theory is defined as
\begin{align}
T_{\mu\sigma}=\langle (\partial_\mu \varphi_a)(\partial_\sigma \varphi_a) - g_{\mu\sigma} \mathcal{L}[\varphi] \rangle\,.
\end{align}
The transverse and longitudinal pressures are given by 
\begin{align}
P_T=\frac{1}{V}\int_{\mathbf{x_T},\eta}\frac{T_{11}+T_{22}}{2}\;, \quad P_L=\frac{1}{\tau^2 V}\int_{\mathbf{x_T},\eta}T_{\eta\eta}\;,
\end{align}
and the energy density is defined to be $\varepsilon = \int_{\mathbf{x_T},\eta}T_{\tau\tau}/V$.  Here we have used the abbreviations  $\int_{\mathbf{x_T},\eta} = \int d^2x_T\,d\eta$ and $V= V_T\, L_\eta$. 

\begin{figure}[tp!]						
 \centering
 \includegraphics[width=0.5\textwidth]{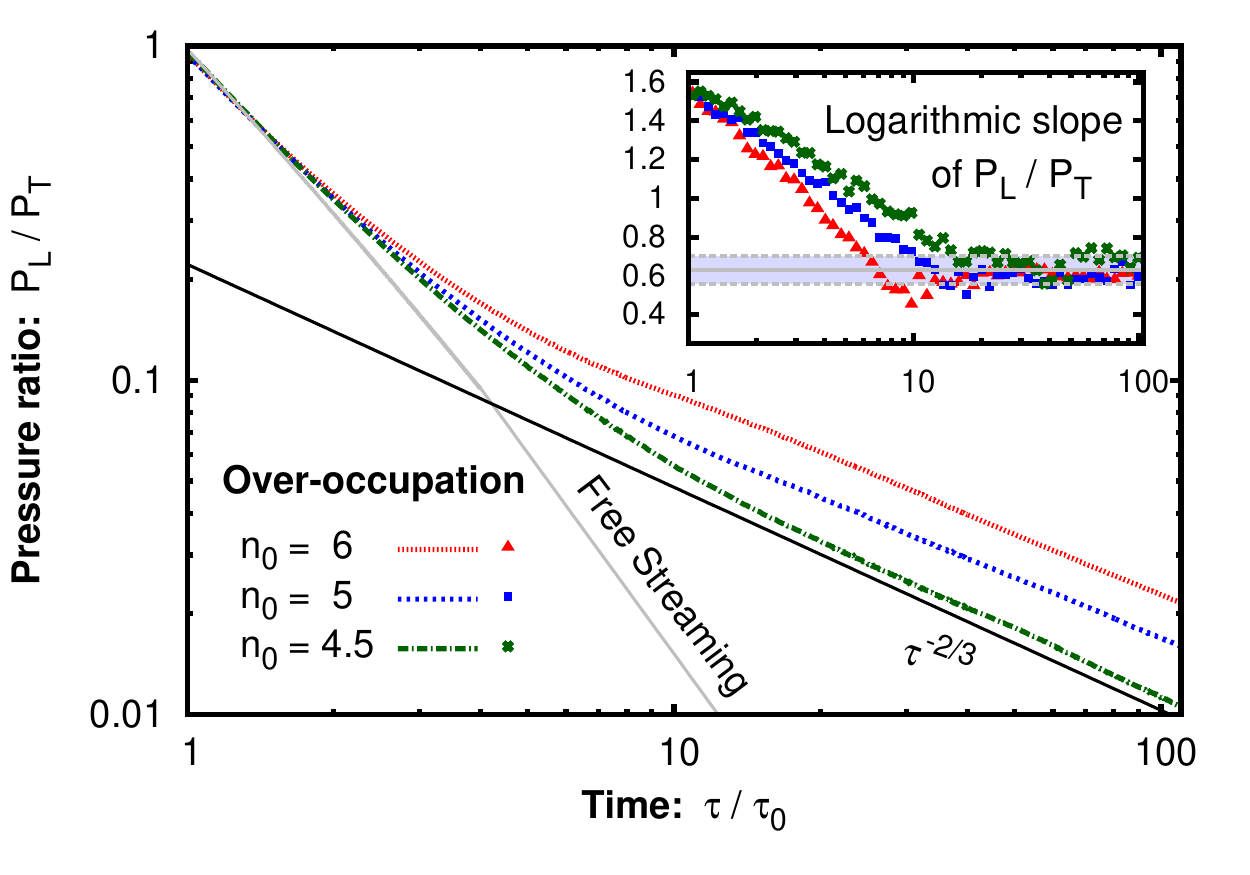}
   \caption{Ratio of the longitudinal to transverse pressure for initial conditions with small initial occupancies $n_{0}$. The ratio $P_L/P_T$ exhibits an approximate $\tau^{-2/3}$ scaling as expected from kinetic theory. In contrast, for the larger occupancies shown in Fig.~\ref{fig:BulkanisotropyGaugeScalar}, a much slower decay of $P_L/P_T$ is observed just as in the case of non-Abelian plasmas.}
  \label{fig:Bulk-anisotropy}
\end{figure}

\begin{figure}[tp!]						
 \centering
 \includegraphics[width=0.5\textwidth]{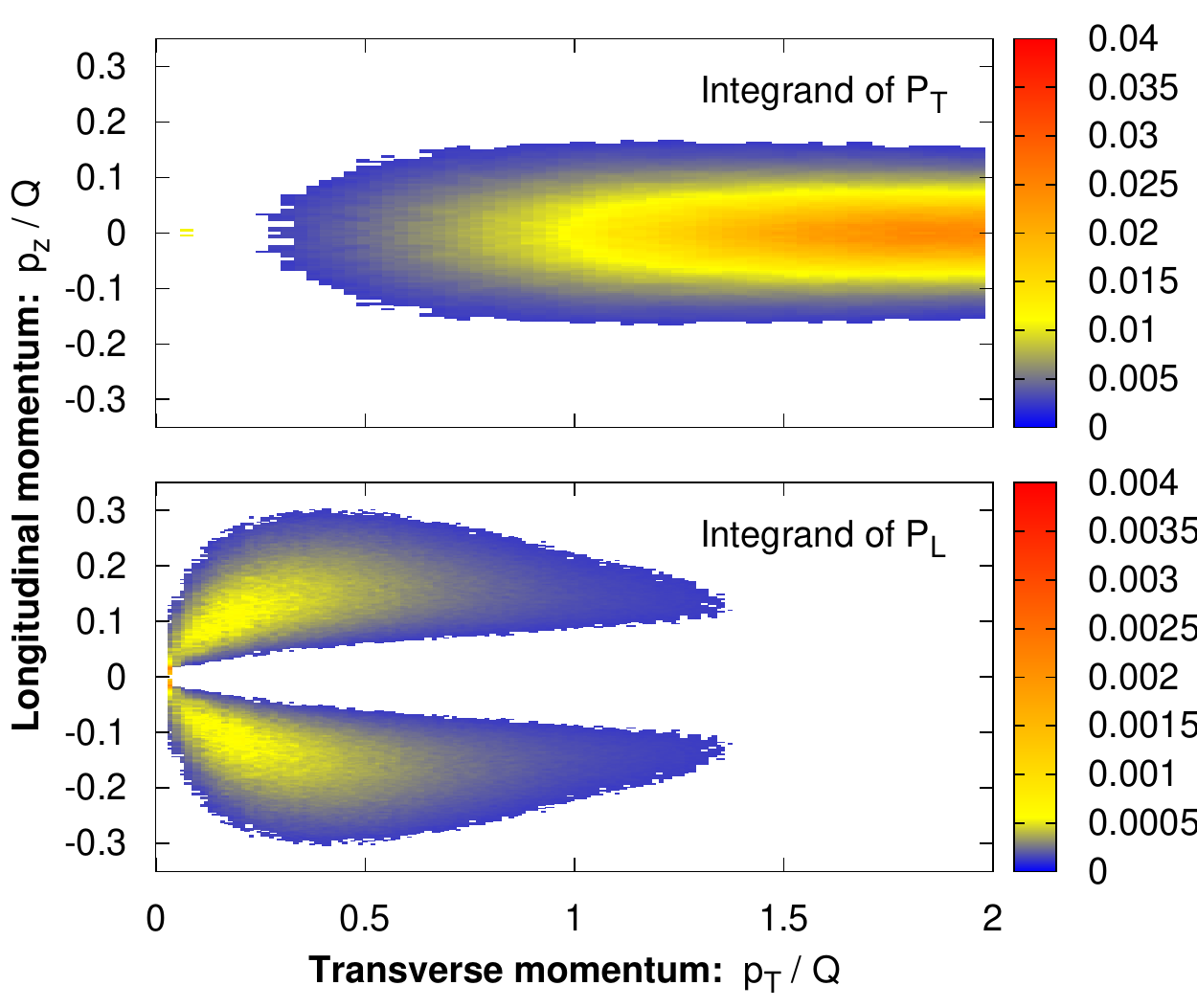}
  \caption{Contour plot of the contributions (see Eq.~(\ref{eq:PressureContributions}))  to the transverse (top) and longitudinal pressure (bottom) for initial occupancy parameter $n_0 = 35$ at time $\tau/\tau_0 = 50$. While the transverse pressure is dominated by hard excitations, one observes that the longitudinal pressure receives its dominant contributions from intermediate and soft momenta.}
  \label{fig:long-pressure-integrand}
\end{figure}

Before we turn to a more detailed discussion of our results for the ratio $P_L/P_T$ presented in Figs.~\ref{fig:BulkanisotropyGaugeScalar} and~\ref{fig:Bulk-anisotropy}, it is useful to first consider the quasiparticle expression for the energy momentum tensor
\begin{align}
T_{\mu\sigma}(\tau)\simeq N \int \frac{d^2p_T\,d p_z}{(2\pi)^3}\, \frac{p_{\mu}p_{\sigma}\; f(p_T,p_z,\tau)}{\omega(p_T,p_z,\tau)}\;,
\label{mat:e-p-tensor-perturbative}
\end{align}
with $\mu,\sigma = x^1, x^2, \eta$ and with the rapidity wave number $p_\eta \equiv \nu = \tau p_z$. We note that for a proper treatment of the nonperturbative dynamics at low momenta, further contributions should be taken into account. While the energy density and the transverse pressure are generally dominated by the hard excitations of the system, the longitudinal pressure in particular can receive significant contributions from the soft sector. We have analyzed this behavior by performing a comparison of the integrands
\begin{align}
\frac{d^2 P_{T}}{dp_Tdp_z}=&\, N \frac{p_T}{(2\pi)^2} \frac{p_T^2\;f(\tau,p_T,p_z)}{2\,\omega(p_T,p_z,\tau)}\;,  \nonumber \\
\frac{d^2 P_{L}}{dp_Tdp_z}=&\, N \frac{p_T}{(2\pi)^2} \frac{p_z^2\;f(\tau,p_T,p_z)}{\omega(p_T,p_z,\tau)}\;.
\label{eq:PressureContributions}
\end{align}
Our results are shown in Fig.~\ref{fig:long-pressure-integrand}, where we show a contour plot of both quantities. While the integrand of the transverse pressure is peaked around transverse momenta of the order of the hard scale, the longitudinal pressure receives its dominant contribution from modes with much softer transverse momenta. This points to the possibility that the quantities $P_L$ and $P_T$ can be dominated by modes from different inertial ranges of momenta.

The time evolution of the pressure ratio $P_L/P_T$ is shown in Figs.~\ref{fig:Bulk-anisotropy} (and earlier in Fig.~\ref{fig:BulkanisotropyGaugeScalar}) as a function of time. While in all cases the bulk anisotropy of the system increases as a function of time, the quantitative behavior on the time scales observed can be different depending on the initial conditions. For low enough initial amplitudes $n_0 \lesssim 6$ shown in Fig.~\ref{fig:Bulk-anisotropy}, one observes an approximate $P_L/P_T \sim \tau^{-2/3}$ scaling.  In this case, a strong infrared enhancement of the spectrum -- corresponding to the scaling regime i) -- is not fully developed during our simulations. Instead the observed behavior is consistent with the expectation that the longitudinal pressure is dominated by modes in the intermediate transverse momentum regime ii). 

In contrast, for the larger values of $n_0$ shown in Fig.~\ref{fig:BulkanisotropyGaugeScalar}, the strong enhancement at soft momenta (given by Eq.~(\ref{mat:fs-low-momenta})) is fully developed and has considerable contributions to the longitudinal pressure. In this case, the combined contributions of soft and intermediate modes lead to a slower decrease of $P_L/P_T$ in time. Interestingly, the approximate scaling behavior observed in this case is quite similar to results obtained for the non-Abelian gauge theory also shown in Fig.~\ref{fig:BulkanisotropyGaugeScalar}. This result is consistent with our earlier observation for the gauge theory case that  modes with relatively small transverse momenta provide a sizeable contribution to the longitudinal pressure (c.f. Ref.~\cite{Berges:2013fga}). 

\section{Range of validity of the classical-statistical approximation}
\label{sec:large-angle}
In this section, we will examine the range of validity of the classical-statistical approximation employed throughout this paper to study the far-from-equilibrium dynamics of a longitudinally expanding scalar theory. Generally the classical-statistical approximation can be justified when the typical occupancies $f$ of the system are large and quantum effects play a negligible role. In particular when the initial occupancies are $f\sim 1/\lambda$ with $\lambda \ll 1$, as considered in this paper, quantum effects are subdominant for times $\tau_{{\rm Quant}} \sim \lambda^{-q}$. Estimating the exponent $q$ requires the evaluation when quantum processes, that are initially suppressed, become important with the increasing dilution of the system.

On the level of kinetic equations, the validity of the classical approximation can be examined by comparing the classical and quantum contributions to the collision integral. One finds that for a perturbative $2\leftrightarrow 2$ scattering process in scalar field theories, the classical part of the collision integral\footnote{At leading order in $1/N$ expansion, the prefactor becomes $\lambda^2/6N$, which is analogous to the prefactor of the vertex-resummed kinetic equation in (\ref{mat:collision-NLO}).} takes the form~\cite{Jeon:2004dh,Micha:2004bv,Berges:2004yj}
\begin{align}
&C_{\text{Classic}}[f](p_1) = \frac{\lambda^2(N+2)}{6N^2} \int d\Omega^{2\leftrightarrow2} \nonumber \\
&\times\Big( (f_{1}+f_{2}) f_{3} f_{4} - f_{1} f_{2} (f_{3}+f_{4}) \Big) \,,
\end{align}
with the corresponding quantum terms
\begin{align}
C_{\text{Quant}}[f](p_1) = \frac{\lambda^2(N+2)}{6N^2} \int d\Omega^{2\leftrightarrow2} \Big( f_{3} f_{4}  - f_{1} f_{2} \Big)\;.
\label{mat:kinetic-quant-corrections}
\end{align}
While the final state Bose enhancement factors are considered to be large in the classical field description, this is not the case when genuine quantum corrections are relevant\footnote{Similar conclusions can be reached also for higher order $n \leftrightarrow m$ processes and in functional nonequilibrium approaches beyond kinetic theory~\cite{Berges:2004yj}.}. One may therefore expect that as long as the classicality condition
\begin{align}
f(p_T,p_z,\tau)\gg 1
\label{mat:classicality-cond}
\end{align}
is satisfied for typical momentum modes, the quantum corrections in Eq.~(\ref{mat:kinetic-quant-corrections}) are suppressed and classical scattering dominates. 

However as recently argued in Ref.~\cite{Epelbaum:2015vxa}, there is potentially a caveat to this argument when the system is highly anisotropic. This is because the classical-statistical approximation restricts the phase space available for each individual scattering to states which are already occupied while the quantum terms allow one to access previously unoccupied states by a single scattering. Assuming that the typical longitudinal momenta $p_z\sim \Lambda_z$ of occupied states are much smaller than the typical transverse momenta $p_T \sim \Lambda_T$ by a factor of $\delta \sim \Lambda_{z}/\Lambda_{T} \ll 1$,  one finds that a classical scattering can only lead to longitudinal momenta on the order of a few times $\sim\Lambda_{z}$. In contrast, the quantum contribution, while suppressed by occupancy factors, has a larger phase space and allows for states with longitudinal momenta up to $\Lambda_{T}\gg\Lambda_{z}$ to be populated by a single scattering. These could in principle give a large contribution to the longitudinal pressure. 

We will show here that this is not the case in the parametric domain of validity of the classical regime. We will begin by first estimating parametrically the importance of `quantum' large angle $2 \leftrightarrow 2$ scatterings into previously unoccupied phase space -- in particular for modes with longitudinal momenta $p_z \sim \Lambda_T \gg \Lambda_z$. We will subsequently investigate whether the dynamics of this process can lead to an earlier breakdown of the classical-statistical approximation than suggested by Eq.~(\ref{mat:classicality-cond}) by estimating the contribution of this process to the longitudinal pressure. 

\subsection{Parametric estimate of quantum large angle scattering}
We begin by estimating the number density of particles at large angles, which can be defined in terms of
\begin{align}
N_{{\rm Large-angle}}(\tau) \sim \int_{|p_{z}|\gg\Lambda_{z}} d^3p ~f(p_T,p_z,\tau)\;,
\end{align}
where the integral over $p_z$ is limited to modes with longitudinal momenta $p_z$ parametrically on the order of $p_T$ -- much bigger than the typical longitudinal momenta $\Lambda_{z}$. We can estimate the change in time of this quantity by considering the Boltzmann equation for $2 \leftrightarrow 2$ scatterings, which, including both classical and quantum contributions, takes the form
\begin{align}
&\partial_{\tau} N_{{\rm Large-angle}}(\tau) \sim  \int_{|p_{1,z}|\gg\Lambda_{z}} d^3 p_{1} \int d^3p_{2} \int d^3p_{3} \int d^3p_{4}  \;  \nonumber \\
&~ \qquad \qquad \frac{\lambda^2}{2\omega_{1} 2\omega_{2} 2\omega_{3} 2\omega_{4}} \delta^{(4)}(p_{1}+p_{2}-p_{3}-p_{4})\; \times \nonumber \\
&~\qquad \Big[ (1+f_{1})(1+f_{2}) f_{3}f_{4} - f_{1}f_{2} (1+f_{3})(1+f_{4}) \Big]\,,
\end{align}
with the abbreviations $f_i = f(p_{i,T},p_{i,z}\tau)$ and similar for $\omega_i$. Since we require $p_{1,z}$ to be much larger than the typical longitudinal momentum of all highly occupied states, i.e. $p_{1,z} \sim \Lambda_{T} \gg \Lambda_{z}$ , longitudinal momentum conservation dictates that also another longitudinal momentum is large, $p_{2,z} \sim \Lambda_{T} \gg \Lambda_{z}$, meaning that both particles fall into the phase space with $f_{1},f_{2} \ll 1$. We can thus neglect the Bose stimulation factors in the gain term (namely, $(1+f_{1})(1+f_{2})\sim 1$) such that only quantum processes contribute to occupancy at large angles. Similarly, we can neglect the loss term altogether since $f_{1}f_{2} \ll 1$, such that our estimate becomes 
\begin{align}
&\partial_{\tau} N_{\rm{Large-angle}}(\tau) \sim \int_{|p_{1,z}|\gg\Lambda_{z}}  dp_{1,z} \int d^2p_{1,T} \int d^3p_{2}\; \times  \nonumber \\ 
&\int d^3p_{3} \int d^3p_{4} ~ \frac{\lambda^2}{2\omega_{1} 2\omega_{2} 2\omega_{3} 2\omega_{4}} \delta^{(4)}(p_{1}+p_{2}-p_{3}-p_{4})~f_{3}f_{4} \,.
\end{align}
We expect the dominant phase space for this scattering to be when the energy of all particles is on the order of the hard transverse scale $\Lambda_{T}$;  therefore parametrically the energy denominator $\omega_{i}\sim\Lambda_T$. One thus obtains the phase space integral for this scattering to be 
\begin{align}
&\partial_{\tau} N_{{\rm Large-angle}}(\tau) \sim \frac{\lambda^2}{\Lambda_T^4}\int_{|p_{1,z}|\gg\Lambda_{z}}  dp_{1,z} \int d^2p_{2,T} \; \times \nonumber \\ 
& \int d^3p_{2} \int d^3p_{3} \int d^3p_{4} ~\delta^{(4)}(p_{1}+p_{2}-p_{3}-p_{4}) ~ f_{3}f_{4}\,.
\end{align}
While at first sight it may seem that this phase space is strongly enhanced for the large angle process, a subtle point is that longitudinal momentum conservation requires $p_{1,z}+p_{2,z} = p_{3,z}+p_{4,z}$ to be on the order of $\sim \Lambda_{z}$, which is the typical longitudinal momentum of highly occupied particles. Implementing this momentum conservation constraint leads to
\begin{align}
&\partial_{\tau} N_{{\rm Large-angle}}(\tau) \sim \frac{\lambda^2}{\Lambda_T^4} \int d^3(p_{1}-p_{2})  \nonumber \\
&~~\int d^3p_{3} \int d^3p_{4} ~\delta(\omega_{1}+\omega_{2}-\omega_{3}-\omega_{4}) ~ f_{3}f_{4}\,.
\end{align}
Since the typical energy of all participating scatterers is of the order of $\Lambda_T$ and there are no further constraints beyond energy conservation, the phase space for the scattering should be of order $\Lambda_T^2$. This leads us to the final estimate
\begin{align}
\partial_{\tau} N_{{\rm Large-angle}}(\tau) \sim \frac{\lambda^2}{\Lambda_{T}^2} N_{{\rm hard}}^2(\tau)\;.
\label{mat:large-angle-rate}
\end{align}
where $N_{{\rm hard}}(\tau)$ denotes the typical density of hard particles
\begin{align}
N_{{\rm hard}}(\tau)\sim \int d^3p~f(p_z,p_T,\tau)\,.
\end{align}
Therefore the total number of particles found at large angles should be of the order of the scattering rate in Eq.~(\ref{mat:large-angle-rate}) times a time scale on the order of the age of the system\footnote{An alternative way to arrive at the same parametric estimate is -- following Ref.~\cite{Baier:2000sb} -- to consider the rate of large angle scatterings that a single hard particle undergoes in the medium. Since the process is not Bose enhanced, this large angle scattering rate is simply given by the density of scattering centers $N_{hard}$ times the cross section
\begin{align}
\frac{dN_{{\rm Large-angle}}^{{\rm Coll}}}{d\tau} \sim \sigma_{{\rm Large-angle}} N_{{\rm hard}}(\tau)\;. \nonumber
\end{align}
The cross section for a large angle scattering of hard scalar particles is parametrically $\sigma_{{\rm Large-angle}}\sim \lambda^2/\Lambda^2_{T}$. The total number of particles at large angles therefore becomes
\begin{align}
N_{{\rm Large-angle}}(\tau) \sim N_{{\rm hard}}(\tau)  \frac{dN_{{\rm Large-angle}}^{{\rm Coll}}}{d\tau} \tau \sim  \frac{\lambda^2}{\Lambda_{T}^2} N_{{\rm hard}}^2(\tau)\,\tau \nonumber
\end{align}
which agrees with the more detailed estimate in the text.}
\begin{align}
N_{Large-angle}(\tau) \sim \lambda^2 \frac{N_{hard}^2(\tau)}{\Lambda_T^3}~\Lambda_{T}\, \tau\;.
\end{align}

\subsection{`Quantum' contributions to observables}
\begin{figure}[tp!]						
 \centering
 \includegraphics[width=0.5\textwidth]{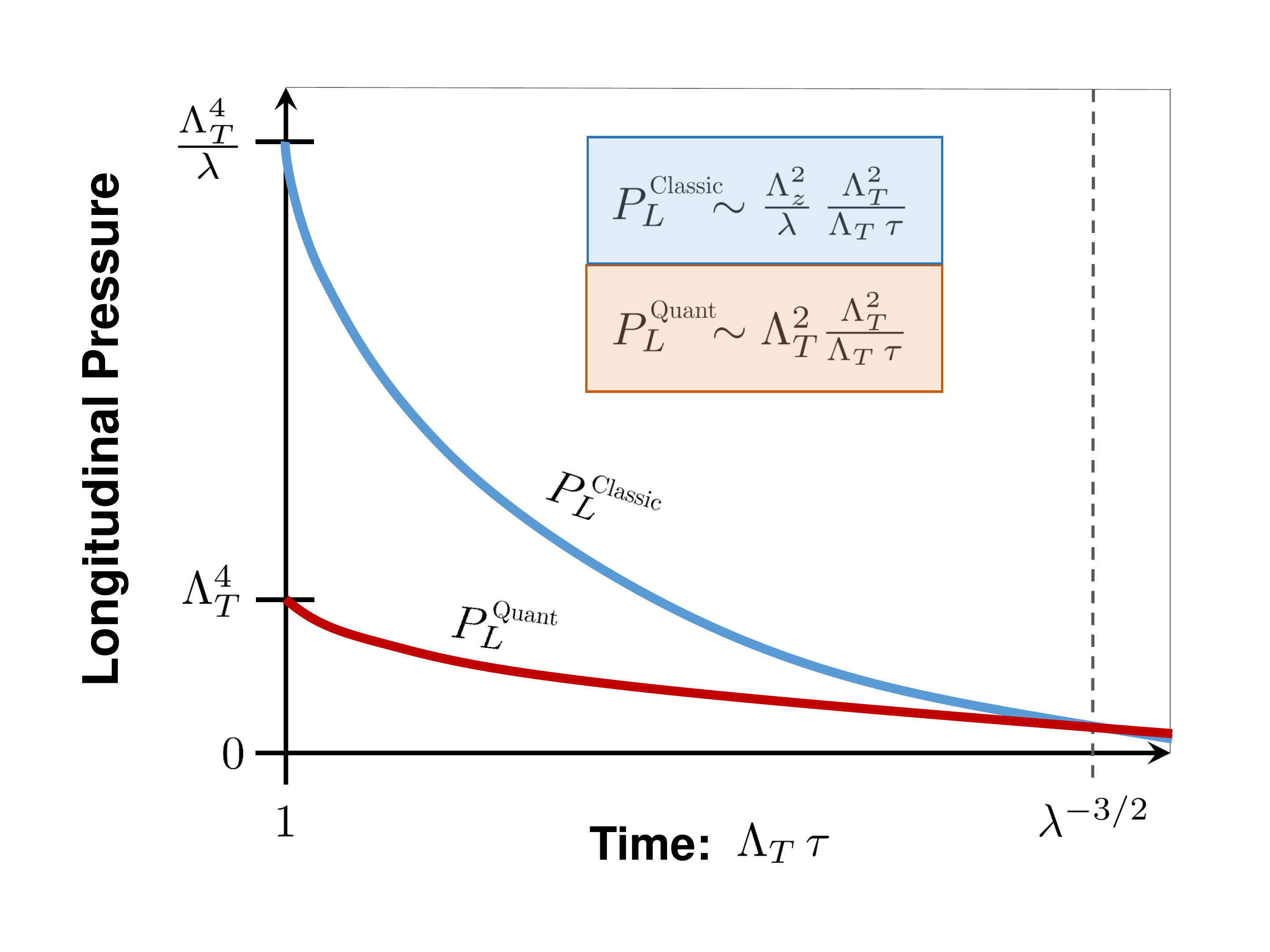}
  \caption{Shown are the small angle and large-angle contributions to the longitudinal pressure as described in the text. While the large-angle contribution is suppressed initially by $1/\lambda$, it becomes important when $\Lambda_z^2 \sim \lambda\,\Lambda_T^2$, which occurs after the time $\lambda^{-3/2}$.}
  \label{fig:long-pressure-small-vs-large-angle}
\end{figure}

With our parametric estimate of the density of large angle particles in hand, we will now estimate the contribution of such quantum effects to the longitudinal pressure\footnote{In general, the relative importance of the leading classical and subleading quantum contributions (as well as other higher order effects) depends on the observable at hand and needs to be evaluated for each observable. While certain ultraviolet sensitive observables may receive large corrections from higher order processes, the dynamics of the bulk may be perfectly well described by the leading order classical process.}. Since the typical longitudinal momenta of the large angle particles are on the order of the transverse hard scale $\Lambda_T$, the contribution to the pressure takes the form
\begin{align}
P_L^{{\rm Quant}}(\tau) \sim N_{{\rm Large-angle}}(\tau) \Lambda_T \sim \lambda^2 \frac{N_{{\rm hard}}^2(\tau)}{\Lambda_T^2}~\Lambda_{T}\, \tau .
\end{align}
which should be compared to the (perturbative) classical contribution
\begin{align}
P_L^{{\rm Classic}}(\tau) \sim N_{{\rm hard}}(\tau) \frac{\Lambda_z^2(\tau)}{\Lambda_T}\,.
\end{align}
Since the total number of hard particles is approximately conserved during the evolution, their density decreases as $\sim 1/\tau$ due to the longitudinal expansion, such that
\begin{align}
N_{{\rm hard}}(\tau) \sim  \left( \frac{\tau_{0}}{\tau}\right) N_{{\rm hard}}(\tau_0) \sim \frac{1}{\lambda}  \frac{\Lambda_{T}^3}{\Lambda_T\; \tau}\;,
\end{align}
and the ratio of the quantum contribution to the pressure over the classical contribution is given by
\begin{align}
\frac{P_L^{\rm Quant}}{P_L^{\rm Classic}} \sim \lambda \frac{\Lambda_T^2}{\Lambda_z^2(\tau)}\;.
\end{align}
At early times\footnote{Even if the initial conditions are such that the typical $p_z\sim 0$ at the instant of the collision, modes at the hard scale are very quickly populated by the instability resulting from the parametric resonance. As noted previously, the characteristic time scale for this is $\tau=\tau_0\ln^{3/2}(\lambda^{-1})$. How rapidly this occurs is clearly visible from Fig.~\ref{fig:coherent-field-IC}. A similar argument applies for gauge theories~\cite{Berges:2013fga,Berges:2014yta}.} $\tau\sim \tau_0$, $\Lambda_z^2 \sim \Lambda_T^2$, hence $P_L^{\rm Quant}/P_L^{\rm Classic} \sim \lambda$, which is very small for $\lambda \ll 1$. With increasing time though the $\Lambda_z^2$ will decrease relative to $\Lambda_T^2$ as illustrated in Fig.~\ref{fig:long-pressure-small-vs-large-angle}. Since on large time scales the longitudinal pressure is dominated by modes in the intermediate rather than hard momentum region, we will assume for this estimate {\it a la} BMSS that
\begin{align}
\Lambda_{z} \sim \Lambda_T (\Lambda_T\tau)^{-1/3} \,.
\end{align}
One then finds that the quantum contribution to the pressure becomes of the same order of magnitude as the classical contribution on the time scale 
\begin{align}
\Lambda_T\,\tau_{{\rm Quant}} \sim \lambda^{-3/2}\,.
\label{mat:tau-quant}
\end{align}
Interestingly however since $f(p_T, p_z \sim \Lambda_{z})\sim (\Lambda_T\tau)^{-2/3}$ in this intermediate transverse momentum regime -- the time scale obtained from Eq.~(\ref{mat:tau-quant}) is parametrically the same as that when the occupancies in this regime become of order unity. We may therefore conclude that the `large angle' quantum contribution to the pressure becomes of the order of the classical contribution only on the time scale when the classical-statistical approximation is breaking down anyway. This is indeed therefore a self-consistent criterion for when classical dynamics can be trusted. Beyond the time scale $\tau_{{\rm Quant}}$, the evolution of the system is expected to change qualitatively. While for non-Abelian gauge theories there has been a lot of recent progress in understanding the dynamics of the quantum regime in more detail~\cite{Kurkela:2014tea,Kurkela:2015qoa}, a similar analysis for the scalar theory is presently not available -- we will return to this point in the future.

\section{Summary}
\label{sec:conclusion}

In this paper, we reported on the results of classical-statistical simulations of the far from equilibrium dynamics of a highly occupied longitudinally expanding $O(N)$ scalar theory. This theory features remarkably rich dynamics which is described simultaneously by three distinct inertial scaling regimes i) at soft, ii) intermediate and iii) at hard transverse momenta. The soft momentum region is governed by an inverse particle cascade feeding a Bose  condensate while the intermediate and hard momentum regions describe longitudinal energy and particle transport. Within each inertial range of momenta, the single particle distribution exhibits a self-similar behavior characterized by universal scaling exponents and scaling functions. We extracted these quantities for each of the momentum sectors; these are briefly summarized in the compilation shown in Table~\ref{tab:fixed-points-table}. We also showed that both initial conditions containing either a coherent field or one with overoccupied particle modes arrived at the same nonthermal fixed points in each of the inertial ranges of momenta.

\begin{table}[t!]						
 \centering
 {\renewcommand{\arraystretch}{1.5}
 \renewcommand{\tabcolsep}{0.2cm}
 \begin{tabular}{||c||c|c|c|c||}
 \hline \hline
 Range & $\alpha$ & $\beta$ & $\gamma$ & $\lambda\,f_S(p_T,p_z)$ \\
 \hline \hline
 i) & $1$ & $2/3$ & $2/3$ & $A_{i}\;((|\mathbf{p}|/b)^{1/2} + (|\mathbf{p}|/b)^{5})^{-1}$ \\
 \hline
 ii) & $-2/3$ & $0$ & $1/3$ & $A_{ii}\;\exp(-p_z^2/(2\sigma_z^2))/p_T$ \\
 \hline
 iii) & $-1/2$ & $0$ & $1/2$ & $A_{iii}\;\text{sech}(p_z/\sigma_z)$ \\
 \hline \hline
 \end{tabular}}
  \caption{Summary of the scaling properties of the three inertial ranges observed in the longitudinally expanding scalar theory, see Sec.~\ref{sec:dynamical-attractor} for details. The amplitudes $A_i$ and $A_{ii}$ are non-universal constants, $A_{iii}$ is constant for $p_T \lesssim 2\,Q$ and then rapidly decreases, and the longitudinal scale $\sigma_z$ can depend on $p_T$.}
  \label{tab:fixed-points-table}
\end{table}

We found that the spectrum in soft regime i) evolves isotropically even though the system is expanding longitudinally. The single particle distribution consists of a strong $\sim |\mathbf{p}|^{-5}$ power law enhancement. We showed that the distribution function in the IR follows a self-similar evolution.  We demonstrated that our numerical results from the classical-statistical simulations are consistent with a vertex-resummed kinetic theory~\cite{Berges:2010ez,Orioli:2015dxa} derived from a resummation of 2PI diagrams at NLO in an $1/N$ expansion~\cite{Berges:2001fi,Aarts:2002dj}. Our analysis suggests that the IR dynamics can be understood in terms of an inverse particle cascade from larger to lower momenta that leads to the formation of a Bose condensate. We found that the condensate is formed with a power law $\tau^{2/3} F(p=0) \sim \tau^{1/\alpha}$ with the exponent $\alpha \simeq 1$ in the low momentum region for sufficiently late times and large volumes. The condensation time diverges with volume parametrically as $\tau_c \sim V$. Once the condensate is formed, it is subject to a $F(p=0) \sim \tau^{-2/3}$ power law decay. We  established the presence of a dynamically generated mass, that decreases with time as $m(\tau) \sim \tau^{-1/3}$. We found that the dynamics of the soft regime i) occurs for $p_{\rm soft} \leq m(\tau)$; this indicates that the IR dynamics is that of a nonrelativistic system. 

Self-interacting scalar field theories in static geometries have IR dynamics reminiscent of turbulent nonrelativistic superfluids in 3+1-dimensions, where relativistic and nonrelativistic scalar field theories share a common universality class~\cite{Orioli:2015dxa}. The generic features of these results are also found for the longitudinally expanding system but differ from the static box results in the values of the dynamical exponents.

We extended our previous analysis of the intermediate momentum regime ii) \cite{Berges:2014bba} and verified the robustness of the nonthermal anisotropic fixed point by varying the initial conditions. For a wide range of initial conditions, the universal temporal scaling exponents of the longitudinal and transverse hard scales as well as the exponent characterizing the overall decrease of the occupancy are identical to those previously studied in depth in the longitudinally expanding non-Abelian gauge theory~\cite{Berges:2013fga}. The functional form of the longitudinal momentum distribution of the nonthermal fixed point of the scalar and gauge theory distributions is a Gaussian distribution and the 2-dimensional transverse momentum follows a thermal-like $p_T^{-1}$ behavior. This universality between the expanding scalar and gauge theories is truly remarkable given the very different scattering matrix elements in the two cases. Understanding this universal behavior in a microscopic framework remains a challenging problem.  

We discussed the emergence of a further scaling regime iii) in the expanding scalar theory, which appears for an inertial range of hard momenta $p_T \gtrsim Q$. It is characterized by a flattening of the single particle transverse momentum distribution at vanishing $p_z$ and a hyperbolic secant distribution along the longitudinal direction. The temporal evolution of the longitudinal momentum scale in this region of hard transverse momenta goes as $p_{z,\rm{typ}} \sim \tau^{-1/2}$, which is faster than the $\sim \tau^{-1/3}$ scaling in regime ii) but distinctly slower than the $\sim \tau^{-1}$ scaling expected if the system were to free stream.  While we confirmed that this behavior corresponds to a scaling regime for a range of initial conditions, we are not aware of a known analog of this scaling in any other systems. Unfortunately extending the gauge theory simulations to the values of $\tau/\tau_0$ where this novel regime appears in the scalar theory simulations is not feasible with present computational resources. 

We also studied the implications of the fixed point structure for the behavior of the dynamically generated mass and the longitudinal and transverse pressures in the expanding scalar theory. We find that the dominant contributions to the effective mass come from the strongly enhanced infrared region and from the condensate. In contrast, in thermal equilibrium hard contributions at the temperature scale dominate the effective (Debye) mass due to the absence of a strong enhancement at low momenta. Similarly, soft contributions are important for the longitudinal pressure. We find that for low occupancies, where a strong infrared enhancement of the spectrum is not fully developed, the longitudinal pressure scales as $\tau^{-5/3}$, as expected in kinetic theory from the intermediate region ii). In contrast, when the occupancies are large enough to generate the steep low momentum distribution associated with the formation of a condensate, we observe that the longitudinal pressure is distinctly different from the expectations of kinetic theory. This is suggestive of a strong contribution of the infrared sector to the longitudinal pressure. A similar behavior of the longitudinal pressure was seen also in the gauge theory case~\cite{Berges:2013fga}; it would be interesting to further explore the consequences of this finding for the validity of hydrodynamic descriptions.

Our results therefore present a challenge for kinetic descriptions. It has been suggested that the dominance of small angle scattering in the expanding scalar theory may be because our classical-statistical simulations miss quantum effects that are argued to be unsuppressed for anisotropic systems even in the classical regime. These arguments further suggest that the quantum effects will give large contributions to the longitudinal pressure that are missed in the classical simulations. We showed however that such quantum contributions to the longitudinal pressure become important precisely when the classical regime ends and the classical-statistical description is no longer valid. Hence our results are robust for  $f\gtrsim 1$ . 

A first principles description of the expanding scalar theory in weak coupling is desirable and should be feasible. Our hope is that the universality of this dynamics with that of expanding non-Abelian plasmas will provide fundamental insight into the microscopic dynamics governing the latter.


\begin{acknowledgments}
We thank J-P.~Blaizot, D.~Gelfand, V.~Kasper, A.~Kurkela, G.~D.~Moore, A.~Pi\~neiro Orioli, B.~Schenke, Q.~Wang and L.~Yaffe for discussions and work related to this project. 

K.B.~thanks HGS-HIRe for FAIR  and HGSFP for support, and Brookhaven National Laboratory and its Nuclear Theory group for hospitality during part of this work. R.V.~thanks Heidelberg University for hospitality and support as an Excellence Professor, and the ExtreMe Matter Institute EMMI for support as an EMMI Visiting Professor. K.B., S.S. and R.V. thank the Institute for Nuclear Theory at the University of Washington for its hospitality and the Department of Energy for partial support during the completion of this work.

This work was supported in part by the German Research Foundation (DFG). S.S.~and R.V.~are supported by US Department of Energy under DOE Contract No.~DE-SC0012704. S.S.~gratefully acknowledges a Goldhaber Distinguished Fellowship from Brookhaven Science Associates. 

This research used resources of the National Energy Research Scientific Computing Center, which is supported by the Office of Science of the U.S. Department of Energy under Contract No. DE-AC02-05CH11231. This work was also performed on the computational resource bwUniCluster funded by the Ministry of Science, Research and the Arts Baden-Württemberg and the Universities of the State of Baden-Württemberg, Germany, within the framework program bwHPC. We gratefully acknowledge their support.
\end{acknowledgments}

\appendix

\section{Initialization, mode functions and distribution function}
\label{sec:mode-functions}

Within the classical-statistical description the initial conditions in Eqs.~(\ref{mat:fluct-IC}) and (\ref{mat:cond-IC}) can be implemented by sampling initial field configurations according to\footnote{When considering fluctuation dominated initial conditions in Eq.~(\ref{mat:fluct-IC}), we drop the quantum $1/2$ in (\ref{mat:class-field-init}).}
\begin{align}
 \varphi_a(\xt,&\eta,\tau_0) = \phi_a(\tau_0) + \int_{\mathbf{p_T},\nu} \sqrt{f(p_T,p_z,\tau_0) + \frac{1}{2}} \nonumber \\
 \times & \left( c_{a,\mathbf{p_T},\nu} ~\xi(p_T,\nu,\tau_0) ~e^{i(\mathbf{p_T} \mathbf{x_T} + \nu \eta)} + c.c. \right)\;,
 \label{mat:class-field-init}
\end{align}
with the abbreviations $\int_{\mathbf{p_T},\nu} = \int d^2p_T\,d\nu/(2\pi)^3$ and $p_z = \nu/\tau$. Here the coefficients $c_{a,\mathbf{p_T},\nu}$ are Gaussian complex random numbers satisfying the relation $\langle c_{a,\mathbf{p_T},\nu}~c^{*}_{b,\mathbf{p_T}',\nu'} \rangle = (2 \pi)^3 \delta_{ab} \delta(\mathbf{p_T} - \mathbf{p_T}') \delta (\nu - \nu')$ while other correlations vanish.  

Equation~(\ref{mat:class-field-init}) is a mode function decomposition of the scalar field $\varphi_a(\xt,\eta,\tau_0)$ with mode functions $\xi(p_T,\nu,\tau)$ satisfying the free equation of motion in Fourier space 
\begin{align}
 \left( \partial_\tau^2 + \frac{1}{\tau}\partial_\tau + p_T^2 + \frac{\nu^2}{\tau^2} + m^2(\tau) \right)\xi(p_T,\nu,\tau) = 0.
 \label{mat:free-EOM-expanding}
\end{align}
Here we included for completeness a possible dynamically generated mass $m(\tau)$, which we will neglect at initial time. The mode functions are normalized as~\cite{Dusling:2011rz,Berges:2013fga}
\begin{align}
 \xi(p_T,\nu,\tau)~\overleftrightarrow{\partial_\tau} ~\xi^{*}(p_T,\nu,\tau) = \frac{i}{\tau}\;,
 \label{mat:mode-function-norm}
\end{align}
with $a\overleftrightarrow{\partial_\tau}b = a~\partial_\tau b - b~\partial_\tau a$. With Eqs.~\eqref{mat:free-EOM-expanding} and \eqref{mat:mode-function-norm} the positive frequency mode function reads
\begin{align}
 \xi(p_T,\nu,\tau) = \frac{\sqrt{\pi}}{2}~e^{\pi \nu/2} H_{i\nu}^{(2)}(p_T~\tau)\;,
 \label{mat:mode-function}
\end{align}
and its complex conjugate is the negative frequency solution of Eq.~(\ref{mat:free-EOM-expanding}). Here $H_{i\nu}^{(2)}(x)$ denotes the Hankel function, which we compute numerically using the method described in Ref.~\cite{Berges:2013fga}. 

For vanishing transverse momentum and mass, Eq.~(\ref{mat:mode-function}) is ill-defined and instead, the mode function can be chosen as~\cite{Berges:2013fga}
\begin{align}
 \xi(p_T=0,\nu,\tau) = \frac{1}{\sqrt{2 \nu}} \left( \frac{\tau}{\tau_0} \right)^{-i\nu}\,,
 \label{mat:mode-function-zero-pt}
\end{align}
which satisfies the free equation of motion (\ref{mat:free-EOM-expanding}) and the normalization condition (\ref{mat:mode-function-norm}).

Now let us consider later times $\tau > \tau_0$. To compute the distribution function, we can use Eq.~(\ref{mat:distr-func-definition}) with the statistical correlator
\begin{align}
 F_{ab}(p_T,\nu,\tau,\tau') = \langle \varphi_a(p_T,\nu,\tau)\varphi^{*}_b(p_T,\nu,\tau')\rangle_{\rm{conn}}/V
 \label{mat:F-classical}
\end{align}
and its derivatives, where $\varphi_a(p_T,\nu,\tau)$ is the spatial Fourier transform of the classical field and $V=V_T L_\eta$ denotes the volume. Here $\langle ~ \rangle_{\rm{conn}}$ denotes the classical-statistical average of the connected correlation function.

It is useful to get a direct relation between the statistical correlation function $F$ and the distribution function $f$ that needs to be consistent with any definition of $f$ including Eq.~(\ref{mat:distr-func-definition}). For this, we assume that a mode decomposition as in Eq.~(\ref{mat:class-field-init}) of the field is also applicable at later times, while interactions enter the (resummed) kinetic equation for the distribution function $f(p_T,p_z,\tau)$ via collision integrals. We will see in App.~\ref{sec:asymptotics} that this corresponds to a quasiparticle assumption, which is consistent with our considerations. For the statistical correlation function~(\ref{mat:F-classical}) and its derivatives (defined above Eq.~(\ref{mat:distr-func-definition})), the mode decomposition leads to 
\begin{align}
 F(p_T,\nu,\tau) \;&= (2 f(p_T,p_z,\tau) + 1)\;|\xi(p_T,\nu,\tau)|^2 \nonumber \\
 \ddot{F}(p_T,\nu,\tau) \;&= (2 f(p_T,p_z,\tau) + 1)\;|\tau\partial_\tau\xi(p_T,\nu,\tau)|^2\,,
 \label{mat:F-quasiparticle}
\end{align}
and similar for $\dot{F}(p_T,\nu,\tau)$. With the help of Eq.~(\ref{mat:mode-function-norm}) one can easily verify that the relations (\ref{mat:F-quasiparticle}) are consistent with the definition of the distribution function for times $\tau > \tau_0$ given by (\ref{mat:distr-func-definition}). Moreover, it follows that the alternative definition 
\begin{align}
 &f(p_T,p_z,\tau) + \frac{1}{2} \nonumber \\
 =\; &\frac{\tau^2}{V_T L_\eta N} \sum_a \left\langle \left| \xi^{*}(p_T,\nu,\tau)\overleftrightarrow{\partial_\tau}\varphi_a(p_T,\nu,\tau) \right|^2 \right\rangle
 \label{mat:f-def-alternative}
\end{align}
is consistent as well. The alternative definition (\ref{mat:f-def-alternative}) is similar to what was used in the expanding gauge theory in Ref.~\cite{Berges:2013fga}. We checked explicitly that both definitions (\ref{mat:distr-func-definition}) and (\ref{mat:f-def-alternative}) yield the same results at transverse momenta larger than the dynamically generated mass $m$. Below the mass scale, a modified dispersion should be taken into account. We note that our definition of the distribution function in Eq.~(\ref{mat:distr-func-definition}) is independent of the specific form of the mode function $\xi$ and can thus also be used at low momenta.

\section{Asymptotic expressions and dispersion relation}
\label{sec:asymptotics}

In this Appendix we consider the late-time expressions for the mode function $\xi(p_T,\nu,\tau)$ and statistical correlation function $F(p_T,\nu,\tau)$. This provides the relation in Eq.~(\ref{mat:F-f-omega-phi-relation}) for finite momenta, which connects $F$ to the distribution function and to the time-dependent dispersion relation
\begin{align}
 \omega(p_T,p_z,\tau) \equiv \sqrt{m^2(\tau) + p_T^2 + \nu^2/\tau^2}\,,
 \label{mat:omega-time-form}
\end{align}
with (time-dependent) longitudinal momentum $p_z = \nu/\tau$ and an effective mass $m(\tau)$. Similarly, the late-time expressions lead to the estimate for the dispersion relation given by Eq.~(\ref{mat:dispersion-relation}).

At late times, we make the ansatz 
\begin{align}
 \xi(p_T,\nu,\tau) = \frac{1}{\sqrt{2 \omega(p_T,p_z,\tau)\;\tau}} e^{-i \Omega(p_T,p_z,\tau)}
 \label{mat:xi-asymp}
\end{align}
for the mode function with the time dependent function $\Omega(p_T,p_z,\tau) = \int_{\tau_0}^{\tau} d\tau' \omega(p_T,p_z,\tau')$. To verify its validity, we need to show that (\ref{mat:xi-asymp}) satisfies the free equation of motion (\ref{mat:free-EOM-expanding}) and the normalization condition (\ref{mat:mode-function-norm}). The latter can be easily confirmed. Since the derivatives of (\ref{mat:xi-asymp}) follow
\begin{align}
 (\partial_\tau^2 + \tau^{-1} \partial_\tau)\;\xi = -\omega^2\;\xi + \mathcal{O}\left( \frac{\omega}{\tau}\xi\right)\,,
\end{align}
the asymptotic form in Eq.~(\ref{mat:xi-asymp}) is also a solution of the free equation for momenta satisfying $\omega(p_T,p_z,\tau)\,\tau \gg 1$. Since in our case, the mass slowly decreases such that $\omega(p_T,p_z,\tau)\,\tau \geq m(\tau)\,\tau \simeq m(\tau_0)\,\tau_0^{1/3}\,\tau^{2/3}$ grows with time, this condition becomes eventually true for all momenta after a finite time. This confirms that Eq.~(\ref{mat:xi-asymp}) is a late-time asymptotic expression for the mode function $\xi(p_T,\nu,\tau)$. We note that alternatively, we get a similar form as in Eq.~(\ref{mat:xi-asymp}) by use of the tools provided in Refs.~\cite{Falcao:2006,Dunster:1988} when starting from Eq.~(\ref{mat:mode-function}).

With Eq.~(\ref{mat:xi-asymp}), one finds the expressions
\begin{align}
 |\xi(p_T,\nu,\tau)|^2 = \frac{1}{2 \omega(p_T,p_z,\tau)\;\tau}\,,
 \label{mat:xi-abs-asymp}
\end{align}
and
\begin{align}
 \partial_\tau \xi(p_T,\nu,\tau) = -i\omega(p_T,p_z,\tau) \;\xi(p_T,\nu,\tau)\,,
 \label{mat:xi-dot}
\end{align}
where we have dropped a term $\mathcal{O}(\xi/\tau)$ in the last expression. These equations, together with~(\ref{mat:F-quasiparticle}), provide the estimate for the dispersion relation in (\ref{mat:dispersion-relation}) and for the correlation function in (\ref{mat:F-f-omega-phi-relation}) for finite momenta.

\section{Collision integral from vertex-resummed kinetic theory}
\label{sec:collision-integral-low-momenta}

Supplementary to the discussion of the low momentum region in Sec.~\ref{sec:infrared-explanation}, we write here the expression of the vertex-resummed collision integral derived in Refs.~\cite{Berges:2010ez,Orioli:2015dxa}. Following standard techniques~\cite{Micha:2004bv,Orioli:2015dxa}, we determine its temporal scaling properties by use of the scaling ansatz (\ref{mat:scaling}) for the distribution function $f(p_T,p_z,\tau)$. 

The collision integral can be decomposed into
\begin{align}
 C[f](p) = \int d\Omega^{2\leftrightarrow 2}\; \frac{\lambda_{\text{eff}}^2[f]}{6N}\; F^{2\leftrightarrow 2}[f]\,,
 \label{mat:collision-NLO-app}
\end{align}
with the integration measure $\int d\Omega^{2\leftrightarrow 2}$, the effective coupling $\lambda_{\text{eff}}^2[f]$ and with the functional $F^{2\leftrightarrow 2}[f](\{p_i\}) = [(f_{1} + f_{2}) f_{3} f_{4} - f_{1} f_{2} (f_{3}+f_{4})]$. The integration measure of elastic scattering reads
\begin{align} 
& \int d\Omega^{2\leftrightarrow 2} = \int\frac{d^3 p_2}{(2\pi)^3}\frac{d^3 p_3}{(2\pi)^3}\frac{d^3 p_4}{(2\pi)^3}  \nonumber\\
& \times\, (2\pi)^{4}\, \delta^{(3)}(\mathbf{p}_1 + \mathbf{p}_2 - \mathbf{p}_3 - \mathbf{p}_4)\, 
\frac{\delta(\omega_{1} + \omega_{2} - \omega_{3} - \omega_{4})}{2\omega_{1}\; 2\omega_{2}\; 2\omega_{3}\; 2\omega_{4}} \, , 
\label{mat:measureNLO}
\end{align}
with $p_1 \equiv p$ and with dispersion relation abbreviated as $\omega_{i} \equiv \omega(p_{T,i},p_{z,i},\tau)$. The effective coupling can be expressed as
\begin{align}
&\lambda^2_{\rm eff}[f]({\mbf p}_1,{\mbf p}_2,{\mbf p}_3,{\mbf p}_4,\tau) \equiv \frac{\lambda^2}{3}
\left[ v[f](\omega_1 + \omega_2,{\mbf p}_1+{\mbf p}_2,\tau) \right. \nonumber \\
&\left. + \;v[f](\omega_1 - \omega_3,{\mbf p}_1-{\mbf p}_3,\tau) + v[f](\omega_1 - \omega_4,{\mbf p}_1-{\mbf p}_4,\tau) \right]\,,
\label{mat:leffrel}
\end{align}
with the abbreviation
\begin{align}
 v[f](\omega,{\mbf p},\tau) \equiv \frac{1}{| 1 + \Pi^{R}[f] (\omega,{\mbf p},\tau)|^2}\,.
 \label{mat:leffrel-2}
\end{align}
The appearance of the `one loop' retarded self-energy in the denominator, 
\begin{align}
	& \Pi^R[f](\omega,\mathbf{p},\tau) \, = \, \frac{\lambda}{12} \int \frac{\mathrm{d}^3q}{(2\pi)^3}\, \frac{f(\mathbf{p_T}-\mathbf{q_T},p_z-q_z,\tau)}{\omega_{\mbf p-\mbf q}\; \omega_{\mbf q}} \nonumber\\
	& \quad \times \bigg[ \frac{1}{\omega_{\mbf q}+\omega_{\mbf p-\mbf q}-\omega-i\epsilon} + \frac{1}{\omega_{\mbf q}-\omega_{\mbf p-\mbf q}-\omega-i\epsilon}  \nonumber\\
	& \quad \,\, +\frac{1}{\omega_{\mbf q}-\omega_{\mbf p-\mbf q}+\omega+i\epsilon} + \frac{1}{\omega_{\mbf q}+\omega_{\mbf p-\mbf q}+\omega+i\epsilon} \bigg]\,,
\label{mat:pir_onshell_rel}
\end{align}
is the result of a $1/N$ expansion of the 2PI effective action up to NLO and corresponds to a nonperturbative resummation of an infinite number of self-energy diagrams~\cite{Berges:2004yj,Aarts:2002dj}.

The exponent $\mu$ characterizing the scaling behavior of the collision integral $ C[f](p_T,p_z,\tau) = \tau^{\mu} \, C[f_S](\bar{p_T},\bar{p_z})$ in Eq.~(\ref{mat:collision-scaling}), with the abbreviations $\bar{p_T} \equiv \tau^{\beta} p_T$ and $\bar{p_z} \equiv \tau^{\gamma} p_z$, can be determined by use of the scaling ansatz (\ref{mat:scaling}) for the distribution function. The functional $F^{2\leftrightarrow 2}[f]$ in Eq.~(\ref{mat:collision-NLO-app}) scales as
\begin{align}
 F^{2\leftrightarrow 2}[f](\{p_i\}) = \tau^{3\alpha}\;F^{2\leftrightarrow 2}[f_S](\{\bar{p_i}\})\,.
\end{align}

For the integration measure and the effective coupling we need to take into account that the dispersion relation is dominated by the effective mass at low momenta, which leads to Eq.~(\ref{mat:omega-large-mass}). Inserting this into the integration measure~(\ref{mat:measureNLO}), we have $\delta(\omega_{1} + \omega_{2} - \omega_{3} - \omega_{4})\rightarrow \delta(({\mbf p}_1^2+{\mbf p}_2^2-{\mbf p}_3^2-{\mbf p}_4^2)/2m(\tau))$
and $2\omega_{1}\, 2\omega_{2}\, 2\omega_{3}\, 2\omega_{4} \rightarrow 16 m^4(\tau)$ to lowest nonvanishing order. According to the discussion below Eq.~(\ref{mat:omega-large-mass}), we can assume an isotropic scaling behavior of the momenta in the IR sector with $\beta = \gamma$. Then the integration measure scales as
\begin{align}
 \int d\Omega^{2\leftrightarrow 2}(\{p_i\}) \simeq \tau^{-2(2\beta + \gamma) + 2\beta + 3\sigma} \int d\Omega^{2\leftrightarrow 2}(\{\bar{p_i}\})\,,
\end{align}
with the mass scaling as $m(\tau) \sim \tau^{-\sigma}$.

According to Eq.~(\ref{mat:pir_onshell_rel}), the retarded self-energy is strongly enhanced in the highly overoccupied low momentum region with $f \gg 1/\lambda$ and therefore determines the scaling behavior of the effective coupling $\lambda_{\text{eff}}^2[f] \sim \lambda^2\,|\Pi^R|^{-2}[f]$. One then finds
\begin{align}
 \lambda_{\text{eff}}^2[f](\{p_i\}) \simeq \tau^{2(2\beta + \gamma) -2\alpha -4\beta -2\sigma}\lambda_{\text{eff}}^2[f_S](\{\bar{p_i}\})\,.
\end{align}

Collecting all parts, the collision integral scales with the exponent
\begin{align}
 \mu = \alpha - 2\beta + \sigma\,.
 \label{mat:mu-exponent}
\end{align}

\end{document}